\pdfoutput = 1
\documentclass[12pt, draftclsnofoot, onecolumn]{IEEEtran}

\usepackage[T1]{fontenc}
\usepackage{cite}
\usepackage{amsmath}
\usepackage{amssymb}
\usepackage{mathtools}
\usepackage{siunitx}
\DeclareSIUnit\dBm{dBm}
\usepackage{bm}
\usepackage{commath}
\usepackage{enumitem}
\usepackage{adjustbox}
\usepackage{array}
\usepackage{nth}

\usepackage[cmintegrals]{newtxmath}
\usepackage[caption=false,font=footnotesize]{subfig}
\usepackage{url}

\newcommand{\vect}[1]{\bm{#1}}
\newcommand{\trans}{^{\mathrm{T}}}
\newcommand{\herm}{^{\mathrm{H}}}

\DeclareMathOperator{\e}{e}
\DeclareMathOperator{\jj}{j}
\DeclareMathOperator{\diag}{diag}
\DeclareMathOperator{\trace}{tr}
\DeclareMathOperator*{\argmin}{argmin}
\DeclareMathOperator*{\argmax}{argmax}

\DeclareMathOperator{\atan2}{atan2}

\let\originalleft\left
\let\originalright\right
\renewcommand{\left}{\mathopen{}\mathclose\bgroup\originalleft}
\renewcommand{\right}{\aftergroup\egroup\originalright}

\usepackage{tikz}
\usepackage{pgfplots}
\usepackage{pgfplotstable}
\pgfplotsset{compat=1.6}
\usepgfplotslibrary{colormaps, groupplots, patchplots}
\usetikzlibrary{arrows.meta, backgrounds, calc, decorations, decorations.markings, decorations.pathreplacing, math, patterns, pgfplots.polar, shapes}
\pgfdeclarelayer{bg}    
\pgfsetlayers{bg,main}  

\usepackage[acronym, automake]{glossaries-prefix}
\setacronymstyle{long-short}
\newacronym{1D}{1D}{one-dimensional}
\newacronym{2D}{2D}{two-dimensional}
\newacronym{3D}{3D}{three-dimensional}
\newacronym{3GPP}{3GPP}{Third Generation Partnership Project}
\newacronym{5G}{5G}{fifth generation}
\newacronym{ADCGM}{ADCGM}{Alternating Descent Conditional Gradient Method}
\newacronym[longplural=angles of arrival]{AOA}{AOA}{angle of arrival}
\newacronym[longplural=angles of departure]{AOD}{AOD}{angle of departure}
\newacronym{AWGN}{AWGN}{additive white Gaussian noise}
\newacronym{BS}{BS}{base station}
\newacronym{cdf}{cdf}{cumulative distribution function}
\newacronym{CP}{CP}{cyclic prefix}
\newacronym{CRLB}{CRLB}{Cram\'er-Rao lower bound}
\newacronym{DL}{DL}{downlink}
\newacronym{ESPEB}{ESPEB}{expected \glsentrytext{SPEB}}
\newacronym{EXIP}{EXIP}{extended invariance principle}
\newacronym[longplural=Fisher information matrices]{FIM}{FIM}{Fisher information matrix}
\newacronym{IoT}{IoT}{internet of things}
\newacronym{LOS}{LOS}{line-of-sight}
\newacronym{MIMO}{MIMO}{multiple-input multiple-output}
\newacronym{mm-Wave}{mm-Wave}{millimeter-wave}
\newacronym{NLOS}{NLOS}{non-LOS}
\newacronym{OFDM}{OFDM}{orthogonal frequency-division multiplexing}
\newacronym{OTDOA}{OTDOA}{observed time difference of arrival}
\newacronym{pdf}{pdf}{probability density function}
\newacronym{PEB}{PEB}{position error bound}
\newacronym{pmf}{pmf}{probability mass function}
\newacronym{PSD}{PSD}{positive semidefiniteness}
\newacronym{RCRLB}{RCRLB}{root \glsentrytext{CRLB}}
\newacronym{RE}{RE}{resource element}
\newacronym{RMSE}{RMSE}{root mean square error}
\newacronym{RTT}{RTT}{round-trip time}
\newacronym{Rx}{Rx}{receiver}
\newacronym[prefixfirst={a\ }, prefix={an\ }]{SDP}{SDP}{semidefinite program}
\newacronym{SPEB}{SPEB}{squared position error bound}
\newacronym{SNR}{SNR}{signal-to-noise ratio}
\newacronym[longplural=times of arrival]{TOA}{TOA}{time of arrival}
\newacronym{Tx}{Tx}{transmitter}
\newacronym{UE}{UE}{user equipment}
\newacronym{ULA}{ULA}{uniform linear array}
\newacronym{UCA}{UCA}{uniform circular array}
\newacronym{UL}{UL}{uplink}
\newacronym{UTDOA}{UTDOA}{uplink TDOA}
\makeglossaries

\newsavebox\glsscratchboxa
\newsavebox\glsscratchboxb
\newsavebox\glsscratchboxc
\newsavebox\glsscratchboxd

\usepackage{algorithm}
\usepackage{algpseudocode}
\algnewcommand{\Input}[1]{%
	\State \textbf{input:} \parbox[t]{.8\linewidth}{\raggedright #1}
}
\algnewcommand{\Output}[1]{%
	\State \textbf{output:} \parbox[t]{.8\linewidth}{\raggedright #1}
}
\algnewcommand{\Initialize}[1]{%
	\State \textbf{initialize:} \parbox[t]{.8\linewidth}{\raggedright #1}
}
\algnewcommand{\Break}{%
	\State \textbf{break}
}
\algdef{SE}[DOWHILE]{Do}{doWhile}{\algorithmicdo}[1]{\algorithmicwhile\ #1}%

\usepackage[pdfencoding=auto, psdextra]{hyperref}
\usepackage{bookmark}

\title{Power Allocation and Parameter Estimation for Multipath-based 5G Positioning}

\author{Anastasios~Kakkavas,~\IEEEmembership{Student~Member,~IEEE,}
	Henk~Wymeersch,~\IEEEmembership{Senior~Member,~IEEE,}
	Gonzalo~Seco-Granados,~\IEEEmembership{Senior~Member,~IEEE,}
	Mario~H.~Casta\~neda~Garc\'ia,~\IEEEmembership{Member,~IEEE,}
	Richard~A.~Stirling-Gallacher,~\IEEEmembership{Member,~IEEE,}
	and~Josef~A.~Nossek,~\IEEEmembership{Life~Fellow,~IEEE}
	
	\thanks{This work was supported in part by the EU-H2020 project Fifth Generation Communication Automotive Research and Innovation (5GCAR),
	and in part by the ICREA Academia program and the
	Spanish Ministry of
	Science, Innovation and Universities project TEC2017-89925-R.}
	\thanks{A.~Kakkavas is with the Munich Research Center, Huawei Technologies Duesseldorf GmbH, 80992 Munich, Germany, and also with the Department of Electrical and Computer Engineering, Technical University of Munich, 80333 Munich, Germany (e-mail: anastasios.kakkavas@huawei.com).}
	\thanks{H. Wymeersch is with the Department of Electrical Engineering, Chalmers University of Technology, 412 58 Gothenburg, Sweden (email: henkw@chalmers.se).}
	\thanks{G. Seco-Granados is with the Department of Telecommunications and Systems Engineering, Universitat Autonoma de Barcelona, Spain (UAB) (e-mail: gonzalo.seco@uab.cat).}
	\thanks{M.~H.~Casta\~{n}eda~Garc\'ia and R.~A.~Stirling-Gallacher are with the Munich Research Center, Huawei Technologies Duesseldorf GmbH, 80992 Munich, Germany (e-mail: mario.castaneda@huawei.com; richard.sg@huawei.com).} 
	\thanks{J.~A.~Nossek is with the Department of Electrical and Computer Engineering, Technical University of Munich, 80333 Munich, Germany (e-mail: josef.a.nossek@tum.de).}
}

\hypersetup{
	pdfinfo={
		Title={Power Allocation and Parameter Estimation for Multipath-based 5G Positioning},
		Author={Anastasios Kakkavas, Henk Wymeersch, Gonzalo Seco-Granados, Mario H. Casta\~neda Garc\'ia, Richard A. Stirling-Gallacher, Josef A. Nossek}
	}
}

\begin{document}

		
	
	\maketitle
	\tikz[overlay,remember picture]
	{
		
		\node at ($(current page.south west)+(1in,0.3cm)$) [rotate=0, anchor=south west] {\parbox{\textwidth}{\footnotesize \footnotesize This work has been submitted to the IEEE for possible publication. Copyright may be transferred without notice, after which this version may no longer be accessible.}};
	}

	\begin{abstract}
		We consider a single-anchor \acrlong{MIMO} \acrlong{OFDM} system with imperfectly synchronized \gls{Tx} and \gls{Rx} clocks, where the \gls{Rx} estimates its position based on the received reference signals. The \gls{Tx}, having (imperfect) prior knowledge about the \gls{Rx} location and the surrounding geometry, transmits the reference signals based on a set of fixed beams. In this work, we develop strategies for the power allocation among the beams aiming to minimize the expected \acrlong{CRLB} for \gls{Rx} positioning. 
		Additional constraints on the design are included to ensure that the \gls{LOS} path is detected with high probability.
		Furthermore, the effect of clock asynchronism on the resulting allocation strategies is also studied. We also propose a gridless compressed sensing-based position estimation algorithm, which exploits the information on the clock offset provided by non-line-of-sight paths, and show that it is asymptotically efficient. 
	\end{abstract}


	\IEEEpeerreviewmaketitle
	
	\glsresetall
	\section{Introduction}
		With the advent of \gls{5G} mobile networks, positioning has attracted lots of research interest. 
		The large chunks of bandwidth available at \gls{mm-Wave} frequencies, as well as the potentially large number of antennas placed at both sides of the communication link are the main driving forces, not only for very high data rates and massive connectivity~\cite{MRD17, SAH+14}, but also for a drastic improvement of the positioning accuracy of cellular networks~\cite{WWP+19}. Recently, within the \gls{3GPP}, besides positioning techniques already existing in previous generations of cellular networks~\cite{DRL+18}, such as \gls{OTDOA}, \gls{UTDOA}, new techniques have been standardized, including \gls{DL}-\gls{AOD}, \gls{UL}-\gls{AOA} and multi-cell \gls{RTT}~\cite{KSH+19}. In addition, proposals for reporting delay and angular multipath measurements to enable single-anchor positioning have been considered~\cite{TR22.872}. With their enhanced positioning capabilities, \gls{5G} systems aim to accommodate use cases like assisted/autonomous driving~\cite{WSD+17}, augmented reality and industrial \gls{IoT}~\cite{TR22.872}.
		
		Single-anchor localization, that leverages the high temporal and angular resolution
		of mm-Wave multiple-input multiple-output (MIMO) systems, has received increasing attention in recent years, as it has the potential to ease the requirements of multi-anchor hearability and interference management. The fundamental limits of single-anchor positioning have been investigated in~\cite{SGD+15} for \gls{LOS} and~\cite{AZA+18, GGD18, MWB+19, KCS+19} for multipath channels with single-bounce-\gls{NLOS} components. 
		
		The single-anchor localization algorithms in the literature can be classified into two categories: one-shot schemes without tracking~\cite{SGD+18, TKL+19, PBC+19, LSW19, WZS19, RKL+20, FCW+19, FCW+20}, and tracking approaches~\cite{GJW+16, LMH+19, WGK+18, TLK+18, LLO+19, MMB+19, KGK+20}. While the latter mainly focus on positon estimation and tracking given the channel parameter measurements, the former also deal with the estimation of the channel parameters, as done in the present work. A three-stage algorithm for position estimation with a \gls{MIMO}-\gls{OFDM} system was proposed in~\cite{SGD+18}, where in the first stage a compressed sensing-based algorithm is used to obtain coarse estimates of the multipath parameters (number of paths, \glspl{TOA}, \glspl{AOD}, \glspl{AOA} and gains), with the coarse estimates refined in the second stage. In the third stage, the refined estimates are mapped to the \gls{Rx} position and orientation and the scatterer/reflector positions using the \gls{EXIP}. A similar approach is followed in~\cite{TKL+19}, with the main difference lying in the mapping from channel parameters to position parameters, where an iterative Gibbs sampling method is employed. In~\cite{PBC+19} range-free angle-based approaches are developed assuming prior map information. 
		An algorithm for localization and synchronization of cooperating full-duplex agents using a single-anchor is developed in~\cite{LSW19}.
		The authors of~\cite{WZS19} propose a protocol and an accompanying algorithm that enables a single-anchor to  (quasi-)simultaneously receive messages from multiple agents in order to localize them using \gls{TOA} and \gls{AOA} measurements. The proposed approach is verified on an experimental setup.
		A \gls{DL} positioning algorithm for a single-antenna \gls{Rx}, based on \gls{TOA} and \gls{AOD} measurements is proposed in~\cite{FCW+19}. The work is extended in~\cite{FCW+20}, where a two-step process is used, with the coarse parameter estimates obtained in the first step used for adaptation of the \gls{Tx} beamforming matrix in the second step. In~\cite{ZLL19} an iterative position estimation and \gls{Tx} beamforming refinement algorithm is developed.
		
		Similar to~\cite{FCW+20, ZLL19}, many works have considered the use of prior knowledge of the \gls{Rx} position at the \gls{Tx} to design beamformers that improve the \gls{Rx}'s localization accuracy. 
		In~\cite{GWS18} \gls{CRLB}-optimal precoders for tracking the \gls{AOD} and \gls{AOA} of a path were designed, taking the uncertainty about their value into account. 
		In~\cite{KDD+17}, assuming a \gls{LOS} channel and a multicarrier system, beamformers minimizing the \gls{TOA} and \gls{AOA} error bounds were proposed, based on the current estimate of the \gls{Rx} position.
		Using a similar setup, but additionally considering multiple users, the authors of~\cite{KDU+18} designed 
		beamformers maximizing a weighted sum of Fisher information on delay, \gls{AOD} and \gls{AOA}.
		Although in a different context, 
		the algorithms and the conclusions of~\cite{ZZS18, ZZS19} are relevant to our \gls{Tx} beamforming problem. In~\cite{ZZS18, ZZS19}, robust beamformers under angular uncertainty were designed and it is concluded that the \gls{Rx} steering vector and its derivative contain all the localization information. Again in a different but still relevant setup, the authors of~\cite{LSZ+13} and~\cite{SSW17} compute the optimal power allocation among multiple anchors for ranging-based localization by solving a \gls{SDP}.
		In~\cite{KSW+19} it was shown that, when the uncertainty about the \gls{Rx} position is not considered, it is optimal to transmit only on the directions corresponding to the \gls{Tx} array steering vector and its derivative. 
		The power allocation among these two directions minimizing the \gls{SPEB} qas analytically calculated in~\cite{KSW+19}.
		When the \gls{Rx} location uncertainty is taken into account, the optimal power allocation among the beams of a given \gls{Tx} beam codebook was computed to minimize the average or maximum \gls{SPEB}.
		
		In this paper, we extend our work in~\cite{KSW+19}. We consider a single-anchor setup and a sparse multipath channel, which comprises the \gls{LOS} path and a number of single-bounce \gls{NLOS} paths, as multi-bounce paths are considered too weak for reception at \gls{mm-Wave} frequencies~\cite{VB03,RBM+12,MGP+14,MMMD21}.
		The \gls{Tx} has only a coarse prior knowledge of the underlying geometry and in addition, the \gls{Tx}-\gls{Rx} clocks are imperfectly synchronized. We optimize the power allocation on a beam codebook for the multipath channel and examine the effect of imperfect synchronization on the resulting power allocation. Also, we develop a novel position estimation algorithm, which is evaluated for the proposed power allocation strategies. The main contributions of the work can be summarized as follows:
		\begin{itemize}
			\item We propose power allocation strategies on a fixed \gls{Tx} beam codebook with the aim of minimizing the expected positioning error of the \gls{Rx}. The optimal solution and a suboptimal one with lower computational complexity are presented and evaluated.
			\item We develop a two-stage position estimation algorithm. The first stage consists of a gridless channel parameter estimation algorithm, based on \cite{BSR17}. The second stage maps the channel parameter estimates to position parameters.
			The information about the clock offset offered by \gls{NLOS} paths in combination with the \gls{LOS} path is exploited so as to discard false alarms.
		\end{itemize}
		
		The rest of the paper is organized as follows. In Sec.~\ref{sec:system model and assumptions} we present the system model and the assumptions of the work. The theoretical bound on positioning accuracy is briefly discussed in Sec.~\ref{sec:position error bound} and the proposed power allocation methods are presented in Sec.~\ref{sec:beam power allocation optimization}. The position estimation algorithm is introduced in Sec.~\ref{sec:position estimation} and numerical evaluations of the proposed approaches are provided in Sec.~\ref{sec:numerical results}. Sec.~\ref{sec:conclusion} concludes the work. 
		
		\textbf{Notation:} 
		We use bold lowercase for vectors, 
		bold uppercase for matrices, 
		non-bold for scalars and calligraphic letters for sets. 
		Depending on its argument, 
		$|\cdot|$ denotes the absolute value of a scalar, 
		the determinant of a matrix or the cardinality of a set. 
		The transpose, conjugate transpose and $p$-norm of a vector/matrix 
		are denoted by $\left(\cdot\right)\trans$, $\left(\cdot\right)\herm$ 
		and $\|\cdot\|_p$ and the Frobenius norm of a matrix
		is denoted by $\|\cdot\|_{\text{F}}$. 
		$\Re\left\{\cdot\right\}$ and $\Im\left\{\cdot\right\}$ denote the real 
		and imaginary part of a complex number and $\arg(\cdot)$ denotes its phase. 
		The $i$-th element of a vector and the $(i,j)$-th element of a matrix 
		are denoted by $[\cdot]_i$ and $[\cdot]_{i,j}$, respectively. 
		$\vect{I}_n$, $\vect{1}$ and $\vect{0}$ denote the identity matrix of size $n$, and the all-ones  all-zeros matrix of the appropriate size. $\diag(\vect{x})$ denotes the diagonal matrix with the elements of $\vect{x}$ on its diagonal.
		The expectation operator is denoted by $\mathbb{E}[\cdot]$ 
		and the sets of real and complex numbers 
		are denoted by $\mathbb{R}$ and $\mathbb{C}$. 
		A multivariate (circularly symmetric complex) Gaussian distribution 
		with mean $\vect{\mu}$ and covariance matrix $\vect{C}$ is denoted by
		$\mathcal{N}\left(\vect{\mu}, \vect{C}\right)$ ($\mathcal{N}_{\mathbb{C}}\left(\vect{\mu}, \vect{C}\right)$). 
		The Hessian of a function $f(\vect{x})$ is denoted as $D_{\vect{x}}^2 f(\vect{x})$.
		
	\section{System Model and Assumptions}
		\label{sec:system model and assumptions}
		
		\subsection{Geometric Model}
\begin{figure}
			\centering
			\includegraphics[scale=1]{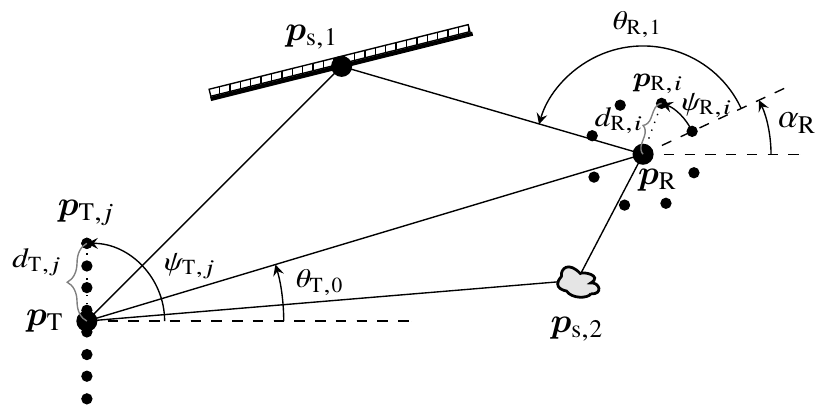}
			\sbox\glsscratchboxa{\footnotesize \gls{ULA}}
			\sbox\glsscratchboxb{\footnotesize \gls{UCA}}
			\sbox\glsscratchboxc{\footnotesize \gls{Tx}}
			\sbox\glsscratchboxd{\footnotesize \gls{Rx}}
			\caption{Geometric model, example with a \unhcopy\glsscratchboxa~at the \unhcopy\glsscratchboxc~and a \unhcopy\glsscratchboxb~at the \unhcopy\glsscratchboxd.}
			\label{fig:geometric_model}
		\end{figure}

		The \gls{Tx} consists of an array with $N_{\text{T}}$ antennas 
		and reference point located at the origin. 
		The \gls{Rx} consists of an array with $N_{\text{R}}$ antennas, 
		a reference point located at 
		$\vect{p}_{\text{R}} = \left[p_{\text{R,x}},\; p_{\text{R,y}}\right]\trans\in\mathbb{R}^2$ 
		and orientation $\alpha_{\text{R}}$. 
		The position of the $j$-th element of the \gls{Tx} array is given by
		\begin{IEEEeqnarray}{rCl}
			\vect{p}_{\text{T},j} &=& d_{\text{T},j}\vect{u}\left(\psi_{\text{T},j}\right)\in\mathbb{R}^2, \quad j=0,\dots, N_{\text{T}} - 1,
			\IEEEeqnarraynumspace 
		\end{IEEEeqnarray}
		where $\vect{u}\left(\psi\right) = \left[\cos\left(\psi\right), \; \sin\left(\psi\right)\right]\trans$ 
		and $d_{\text{T},j}$ and $\psi_{\text{T},j}$ are its distance and angle from the \gls{Tx} array's reference point, 
		as shown in Fig.~\ref{fig:geometric_model}. 
		Accordingly, the position of the $i$-th element of the \gls{Rx} array is given by
		\begin{IEEEeqnarray}{rCl}
			\vect{p}_{\text{R}, i} &=& d_{\text{R}, i} \vect{u}(\psi_{\text{R},i} + \alpha_{\text{R}})\in\mathbb{R}^2, \quad i=0,\dots, N_{\text{R}} - 1.
		\end{IEEEeqnarray}
		
		We assume that for all antenna pairs there are $L$ discrete propagation paths. 
		The first of these $L$ paths ($l=0$) is the \gls{LOS} path 
		and the rest ($l=1,\ldots,L-1$) are single-bounce \gls{NLOS} paths. 
		The point of incidence of the $l$-th single-bounce path, 
		which corresponds either to scattering or reflection, 
		is $\vect{p}_{\text{s},l} = \left[p_{\text{s},l,\text{x}},\; p_{\text{s},l,\text{y}}\right]\trans,\; l=1,\ldots,L-1$. 
		The array apertures are assumed to be small 
		compared to the distance between \gls{Tx} and \gls{Rx}, 
		as well as the distance between each of the scatterers/reflectors and the \gls{Tx} or \gls{Rx}. 
		Therefore, the delay of the $l$-th path from \gls{Tx} element $j$ 
		to \gls{Rx} element $i$ can be approximated by~\cite{KCS+19}
		\begin{IEEEeqnarray}{rCl}
			\tau_{l,i,j} &\approx& \tau_l - \tau_{\text{T},j}(\theta_{\text{T},l}) - \tau_{\text{R},i}(\theta_{\text{R},l}), \quad l=0, \ldots, L-1,
			\label{eq:tau_l,i,j}
			\IEEEeqnarraynumspace
		\end{IEEEeqnarray}
		where 
		\begin{IEEEeqnarray}{rCl}
			\tau_l &=& \begin{cases}
				\|\vect{p}_{\text{R}}\|_2/c + \epsilon_{\text{clk}}, & l=0\\
				\big(\|\vect{p}_{\text{s},l}\|_2 + \|\vect{p}_{\text{R}} - \vect{p}_{\text{s},l}\|_2\big)/c + \epsilon_{\text{clk}}, & l\neq 0,
			\end{cases}\label{eq:pos2ch tau}\\
			\tau_{\text{T},j}(\theta_{\text{T},l}) &=& d_{\text{T},j}\vect{u}\trans(\psi_{\text{T},j})\vect{u}(\theta_{\text{T},l})/c,\\
			\tau_{\text{R},i}(\theta_{\text{R},l}) &=&  d_{\text{R},i}\vect{u}\trans(\psi_{\text{R},i})\vect{u}(\theta_{\text{R},l})/c,
		\end{IEEEeqnarray} 
		with $\epsilon_{\text{clk}}$ being the clock offset between \gls{Tx} and \gls{Rx} 
		and $c$ the speed of light. 
		The angles are defined as 
		\begin{IEEEeqnarray}{rCl}
			\theta_{\text{T},l} &=& \begin{cases}
				\atan2\left(p_{\text{R},\text{y}}, p_{\text{R},\text{x}}\right), & l=0\\
				\atan2\left(p_{\text{s},l,\text{y}}, p_{\text{s},l,\text{x}}\right), & l\neq 0
			\end{cases}\label{eq:pos2ch thT}\\
			\theta_{\text{R},l} &=& \begin{cases}
				\theta_{\text{T},l} + \pi - \alpha_{\text{R}}, & l=0,\\
				\atan2\left(p_{\text{s},l,\text{y}} - p_{\text{R}, \text{y}}, p_{\text{s},l,\text{x}} - p_{\text{R}, \text{x}}\right) - \alpha_{\text{R}}, & l\neq 0,
			\end{cases}\label{eq:pos2ch thR}
		\end{IEEEeqnarray}
		with $\atan2\left(y,x\right)$ being the four-quadrant inverse tangent function.
		
		\subsection{Signal Model}
		An \gls{OFDM} waveform with subcarrier spacing $\Delta f$, $N$ subcarriers
		and \gls{CP} duration $T_{\text{CP}}$ is considered. 
		The reference signal is transmitted on $N_{\text{P}}$ subcarriers, 
		whose indices are described by $\mathcal{P} = \{p_1,\ldots, p_{N_{\text{P}}}\}$
		and $N_{\text{B}}$ \gls{OFDM} symbols are transmitted. 
		We assume a narrowband signal model, 
		i.e. $B/f_{\text{c}} \ll \lambda_{\text{c}}/D_{\max}$, where $B\approx \Delta f (\max(\mathcal{P}) - \min(\mathcal{P}))$ is the signal bandwidth, $f_c$ is the carrier frequency, $\lambda_{\text{c}}$ is the carrier wavelength and $D_{\max}$ is the largest of the \gls{Tx} and \gls{Rx} array apertures.
		The reference signal resource grid $\mathcal{R}$ comprises all resource elements at the time-frequency points $(p,b),\; p\in\mathcal{P},\;b=0,\ldots,N_{\text{B}} - 1$. 
		The transmitter uses a beam codebook $\{\vect{f}_k\}_{k=1}^{M_\text{T}}$,
		where $M_{\text{T}}$ is the number of beams in the codebook and $\|\vect{f}_k\|_2 = 1,  \forall k$. 
		The $k$-th beam is used on a subset $\mathcal{R}_k$ of \glspl{RE} $(p,b)$,
		with $\mathcal{R}_k \cap \mathcal{R}_{k'} = \emptyset$ for $k\neq k'$. 
		The transmitted signal vector at the $p$-th subcarrier, $p\in\mathcal{P}$,
		of the $b$-th OFDM symbol, $b = 0, \ldots, N_{\text{B}} - 1$, then is 
		\begin{IEEEeqnarray}{rCl}
			\vect{x}[p, b] &=& \lambda_k[p, b]\vect{f}_k,\; (p,b)\in\mathcal{R}_k, \label{eq:x[p,b] definition}
		\end{IEEEeqnarray}
		where 
		\begin{IEEEeqnarray}{rCl}
			\lambda_k[p, b] &=& \sqrt{P_{\text{tot}} q_k \gamma_k[p,b]}\e^{\jj\beta_k[p,b]}\label{eq:lambda_k[p, b]} \label{eq:lambda_k[p,b] definition}
		\end{IEEEeqnarray} 
		is the symbol assigned to $\vect{f}_k$ at the $p$-th subcarrier, $P_{\text{tot}}$ is the total \gls{Tx} power (disregarding the power spent for the \gls{CP}),
		$q_k$ is the fraction of $P_{\text{tot}}$ allocated to $\vect{f}_k$, 
		with $\sum_{k=1}^{M_{\text{T}}} q_k = 1$, 
		$\gamma_k[p,b]$ is the fraction of $q_k$ allocated to the RE $(p,b)$, 
		with $\sum_{(p,b)\in\mathcal{R}_k}\gamma_k[p,b] = 1$, 
		and $\beta_k[p,b]$ is the phase of $\lambda_k[p,b]$. 
		The received signal is
		\begin{IEEEeqnarray}{rCl}
			\vect{y}[p, b] &=& \vect{m}[p, b] + \vect{\eta}[p,b],
			\label{eq:signal model}
			\IEEEeqnarraynumspace
		\end{IEEEeqnarray}
		where 
		\begin{IEEEeqnarray}{rCl}
			\vect{m}[p,b] &=& \sum_{l=0}^{L-1} h_l \e^{-\jj \omega_p \tau_l} \vect{a}_{\text{R}}(\theta_{\text{R},l}) \vect{a}_{\text{T}}\trans(\theta_{\text{T},l}) \vect{x}[p,b],\label{eq:m[p,b] definition}\\
			\vect{a}_{\text{T}}(\theta_{\text{T},l}) &=& \begin{bmatrix}
				e^{\jj \omega_c \tau_{\text{T},1}(\theta_{\text{T},l})}, & \ldots, &\e^{\jj \omega_c \tau_{\text{T},N_{\text{T}}}(\theta_{\text{T},l})}
			\end{bmatrix}\trans\in\mathbb{C}^{N_{\text{T}}}
			\IEEEeqnarraynumspace
		\end{IEEEeqnarray}
		is the \gls{Tx} array steering vector, with the \gls{Rx} steering vector $\vect{a}_{\text{R}}(\theta_{\text{R}, l})$ defined accordingly, 
		$\omega_p = 2\pi p \Delta f,\; \omega_{\text{c}} = 2\pi f_{\text{c}}$, $h_l$ is the gain of the $l$-th path 
		and $\vect{\eta}[p,b] \sim \mathcal{N}_{\mathbb{C}}(\vect{0},\sigma_{\eta}^2\vect{I}_{N_{\text{R}}})$ is the \gls{AWGN}. 
		We write the signal model~\eqref{eq:signal model} as
		\begin{IEEEeqnarray}{rCl}
			\vect{Y}_b &=& \sum\nolimits_{l=0}^{L-1} h_l \vect{C}_b(\tau_l, \theta_{\text{T},l}, \theta_{\text{R},l}) + \vect{N}_b,
		\end{IEEEeqnarray}
		where
		\begin{IEEEeqnarray}{rCl}
			\hspace{-0.5cm}\vect{C}_b(\hspace{-0.03cm}\tau_l, \hspace{-0.03cm}\theta_{\text{T},l}, \hspace{-0.03cm}\theta_{\text{R},l}\hspace{-0.03cm}) &=& \vect{a}_{\text{R}}(\hspace{-0.04cm}\theta_{\text{R},l}\hspace{-0.03cm}) \vect{a}_{\text{T}}\trans(\hspace{-0.03cm}\theta_{\text{T},l}\hspace{-0.03cm})\vect{X}_b \hspace{-0.05cm}\diag(\hspace{-0.04cm}\vect{a}_{\tau}(\hspace{-0.02cm}\tau_l\hspace{-0.02cm})\hspace{-0.03cm})\hspace{-0.05cm}\in \hspace{-0.05cm}\mathbb{C}^{N_{\text{R}}\times N_{\text{P}}}\label{eq:C_b definition}\\
			\vect{a}_{\tau}(\tau) &=& [\e^{-\jj\omega_{p_1}\tau},\ldots, \e^{-\jj\omega_{p_{N_{\text{P}}}}\tau}]\trans\in\mathbb{C}^{N_{\text{P}}}\\
			\vect{Y}_b &=& [\vect{y}[p_1, b], \ldots, \vect{y}[p_{N_{\text{P}}}, b]]\in \mathbb{C}^{N_{\text{R}}\times N_{\text{P}}}\\
			\vect{X}_b &=& [\vect{x}[p_1, b], \ldots, \vect{x}[p_{N_{\text{P}}}, b]]\in \mathbb{C}^{N_{\text{T}}\times N_{\text{P}}}\\
			\vect{N}_b &=& [\vect{\eta}[p_1, b], \ldots, \vect{\eta}[p_{N_{\text{P}}}, b]]\in \mathbb{C}^{N_{\text{R}}\times N_{\text{P}}}
			\IEEEeqnarraynumspace
		\end{IEEEeqnarray}
		Stacking the observations over $N_{\text{B}}$ OFDM symbols we get
		\begin{IEEEeqnarray}{rCl}
			\vect{Y} = \sum_{l=0}^{L-1} h_l \vect{C}(\tau_l, \theta_{\text{T},l}, \theta_{\text{R},l}) + \vect{N}\label{eq:all observations stacked}
		\end{IEEEeqnarray}
		where 
		\begin{IEEEeqnarray}{rCl}
			\vect{Y} &=& [\vect{Y}_0\trans,\ldots,\vect{Y}_{N_{\text{B}} - 1}\trans]\trans\\
			\vect{C}(\tau, \theta_{\text{T}}, \theta_{\text{R}}) &=& [\vect{C}_0\trans(\tau, \theta_{\text{T}}, \theta_{\text{R}}),\ldots,\vect{C}_{N_{\text{B}} - 1}\trans(\tau, \theta_{\text{T}}, \theta_{\text{R}})]\trans\\
			\vect{N} &=& [\vect{N}_0\trans,\ldots,\vect{N}_{N_{\text{B}} - 1}\trans].
		\end{IEEEeqnarray}

		Through~\eqref{eq:pos2ch tau},~\eqref{eq:pos2ch thT}-\eqref{eq:pos2ch thR} and \eqref{eq:all observations stacked}, 
		we can see that the observations $\vect{Y}$ depend on the position parameter vector $\vect{\nu}$, defined as
		\begin{IEEEeqnarray}{rCl}
			\vect{\nu} &=&  [\vect{p}_{\text{R}}\trans, \alpha_{\text{R}}, \epsilon_{\text{clk}}, \vect{h}_0\trans, \vect{p}_{\text{s}, 1}\trans, \vect{h}_1\trans,\ldots,\vect{p}_{\text{s}, L-1}\trans, \vect{h}_{L - 1}\trans]\trans\in \mathbb{R}^{4L + 2},
			\IEEEeqnarraynumspace
			\label{eq:position parameter vector definition}
		\end{IEEEeqnarray}
		with $\vect{h}_l = [|h_l|, \arg(h_l)]\trans$.
		
		\subsection{Assumptions}
		\subsubsection{Reference signal structure}
		In this work we consider the case where \gls{Tx} uses a fixed beam codebook $\vect{f}_k,\;k=1,\ldots,M_{\text{T}}$. 
		This does not only simplify the optimization task, 
		but also might be a practical limitation in a \gls{5G} system, 
		with devices using a predefined set of beams for transmission or reception. 
		
		We also assume that the resource allocation $\mathcal{R}_k$ among the codebook beams
		and the power allocation $\gamma_k[p,b]$ among assigned \glspl{RE}, 
		are fixed and therefore, 
		optimizing $\mathcal{R}_k$ is not in the scope of our reference signal optimization task. The problem of designing a waveform the has been addressed in~\cite{DJR+16, LSS11, SSW15}, where the \gls{CRLB} is optimized with respect to the resource allocation and additional constraints may be considered to avoid a unbalanced use of the spectrum.
		
		\subsubsection{Prior knowledge at \gls{Rx} and \gls{Tx}}
		In many cases the \gls{Tx} might have prior knowledge on $\vect{\nu}$,
		based on prior estimation in the reverse link, 
		map information and known geographical distribution of the users. 
		The prior information is encoded by the joint \gls{pdf} $p_{\vect{\nu}}(\vect{\nu})$. 
		In the following, we examine how the \gls{Tx} can expoit the prior information, 
		so as to improve the ability to localize the \gls{Rx}.
		
		The \gls{Rx}, which aims to compute its position and orientation from the received signal, 
		only has knowledge on the clock offset's distribution $p_{\epsilon_{\text{clk}}}$, which we assume to be zero-mean Gaussian with variance $\sigma_{\text{clk}}^2$. 
		We note that $\sigma_{\text{clk}} = 0$ and $\sigma_{\text{clk}} \rightarrow \infty$ 
		correspond to perfect synchronization and asynchronous operation, respectively.

	\section{Position Error Bound}
		\label{sec:position error bound}
		
		The achievable positioning accuracy of the \gls{Rx} can be characterized in terms of the hybrid \gls{CRLB}. 
		For a parameter vector $\vect{\nu}$ containing both deterministic and random paramters, the covariance matrix $\vect{C}$ of any unbiased estimator $\hat{\vect{\nu}}$ of $\vect{\nu}$ satisfies~\cite{RS87,Mess06} 
		\begin{IEEEeqnarray}{rCl}
			\vect{C} - \vect{J}_{\vect{\nu}}^{-1} \succeq \vect{0},
		\end{IEEEeqnarray}
		where $\succeq \vect{0}$ denotes positive semi-definiteness 
		and $\vect{J}_{\vect{\nu}}\in\mathbb{R}^{(4L + 2)\times (4L + 2)}$ is the hybrid \gls{FIM} of $\vect{\nu}$. 
		$\vect{J}_{\vect{\nu}}$ is defined as
		\begin{IEEEeqnarray}{rCl}
			\vect{J}_{\vect{\nu}} = \vect{J}^{(\text{p})}_{\vect{\nu}} + \vect{J}^{(\text{o})}_{\vect{\nu}}\label{eq:J_phi definition},
		\end{IEEEeqnarray}
		where 
		\begin{IEEEeqnarray}{rCl}
			\vect{J}^{(\text{p})}_{\vect{\nu}} &=& \mathbb{E}_{\vect{\nu}_r}[-D_{\vect{\nu}}^2\ln p(\vect{\nu}_r)],
		\end{IEEEeqnarray}
		accounts for the prior information and
		\begin{IEEEeqnarray}{rCl}
			\vect{J}^{(\text{o})}_{\vect{\nu}} &=& \mathbb{E}_{\vect{Y},\vect{\nu}_r}[-D_{\vect{\nu}}^2\ln p(\vect{Y}|\vect{\nu})]
		\end{IEEEeqnarray}
		accounts for the observation-related information, 
		with $\vect{\nu}_r$ representing the random parameters in $\vect{\nu}$.
		As $\epsilon_{\text{clk}}$ is the only parameter with prior information at the \gls{Rx},
		it is straightforward to find that, 
		based on \eqref{eq:position parameter vector definition}, 
		the only non-zero entry of $\vect{J}_{\vect{\nu}}^{(\text{p})}$ is 
		\begin{IEEEeqnarray}{rCl}
			\big[\vect{J}^{(\text{p})}_{\vect{\nu}}\big]_{4,4} &=& 1/\sigma_{\text{clk}}^2. \label{eq:J_clk_prior}
		\end{IEEEeqnarray}
		Since $\vect{\nu}$ is observed under \gls{AWGN}, the $(i,j)$-th entry of the $\vect{J}_{\vect{\nu}}^{(\text{o})}$ is
		\begin{IEEEeqnarray}{rCl}
			\left[\vect{J}_{\vect{\nu}}^{(\text{o})}\right]_{i,j} &=& \frac{2}{\sigma_{\eta}^2}\sum_{b=1}^{N_{\text{B}}} \sum_{p\in \mathcal{P}} \Re\left\{\frac{\partial \vect{m}_b\herm[p]}{\partial \nu_i} \frac{\partial \vect{m}_b[p]}{\partial \nu_j}\right\}.
			\label{eq:entries of channel parameter FIM}
			\IEEEeqnarraynumspace
		\end{IEEEeqnarray}
		Using \eqref{eq:pos2ch tau}, \eqref{eq:m[p,b] definition} and \eqref{eq:entries of channel parameter FIM}, we can see that $\vect{J}_{\vect{\nu}}^{(\text{o})}$ is independent of the value of $\epsilon_{\text{clk}}$.
		The \gls{SPEB} is defined as
		\begin{IEEEeqnarray}{rCl}
			\text{SPEB} = \trace(\vect{E}\trans\vect{J}_{\vect{\nu}}^{-1}\vect{E}),
		\end{IEEEeqnarray}
		where $\vect{E} = [\vect{e}_1,\;\vect{e}_2]$ 
		and $\vect{e}_i$ is the $i$-th column of the identity matrix of the appropriate size. The \gls{PEB} is defined as its square root.


	\section{Beam Power Allocation Optimization}
		\label{sec:beam power allocation optimization}
		
		For the reference signal optimization, we make use of the assumption that with large bandwisth and number of antennas the paths are asymptotically orthogonal \cite{AZA+18, KCS+19}. We note that the \gls{SPEB} is a function of 
		\begin{IEEEeqnarray}{rCl}
			\vect{\nu}' = [\vect{p}_{\text{R}}\trans, \alpha_{\text{R}}, |h_0|, \vect{p}_{\text{s}, 1}\trans, |h_1|,\ldots,\vect{p}_{\text{s}, L-1}\trans, |h_{L-1}|\trans]\trans\in\mathbb{R}^{3L + 1},
			\IEEEeqnarraynumspace
		\end{IEEEeqnarray}
		that is, it is independent of the values of $\arg(h_l),\; l = 1,\ldots, L - 1$, and $\epsilon_{\text{clk}}$. 
		Also, due to the inner product of the derivatives in \eqref{eq:entries of channel parameter FIM}, 
		we can observe (see \eqref{eq:x[p,b] definition}, \eqref{eq:lambda_k[p,b] definition} and \eqref{eq:m[p,b] definition}) 
		that $\vect{J}$ is independent of $\beta_k[p,b]$.
		In the following, we write $\vect{J}_{\vect{\nu}} = \vect{J}_{\vect{\nu}}(\vect{q},\vect{\nu}')$, with $\vect{q} = [q_1,\ldots,q_{M_{\text{T}}}]\in\mathbb{R}^{M_{\text{T}}}$,
		to stress that $\vect{J}_{\vect{\nu}}$ is the hybrid \gls{FIM} of $\vect{\nu}$, 
		whose value depends on $\vect{q}$ and $\vect{\nu}'$.
		Similarly, we write $\text{SPEB} = \text{SPEB}(\vect{q}, \vect{\nu}')$.
		
		We study how the \gls{Tx} can optimize the beam power allocation $\vect{q}$
		using its prior knowledge on $\vect{\nu}'$ 
		so as to enable higher positioning accuracy at the \gls{Rx}. 
		We choose the \gls{ESPEB}
		\begin{IEEEeqnarray}{rCl}
			\text{ESPEB} &=& \mathbb{E}_{\vect{\nu}'}[\text{SPEB}(\vect{q}, \vect{\nu}')]	
		\end{IEEEeqnarray}
		as the performance metric. The following proposed methods can be easily adapted for other objectives, such as $\max_{\vect{\nu}'}\text{SPEB}(\vect{q}, \vect{\nu}')$.
		
		\subsection{Problem formulation}
		The optimization problem in hand reads as:
		\begin{IEEEeqnarray}{rCl}
			\min_{\vect{q}} \mathbb{E}_{\vect{\nu}'}[\text{SPEB}(\vect{q}, \vect{\nu}')]\;\; \text{s.t. } & \vect{q}\succcurlyeq \vect{0},\; \vect{1}\trans\vect{q}\leq 1,
			\label{eq:original problem}
		\end{IEEEeqnarray}
		where $\succcurlyeq$ denotes element-wise inequality.
		In order to solve \eqref{eq:original problem}, 
		one can employ a cubature rule \cite{Coo03, Cro14} with positive weights to approximate the expectation integral with a sum:
		\begin{IEEEeqnarray}{rCl}
			\mathbb{E}_{\vect{\nu}'}[\text{SPEB}(\vect{q}, \vect{\nu}')] &\approx& \sum\nolimits_{j=1}^{N_{\vect{\nu}}'} p_j \text{SPEB}(\vect{q}, \vect{\nu}'_j),
		\end{IEEEeqnarray}
		where $\vect{\nu}_j'$ and $p_j>0, \; j = 1,\ldots, N_{\vect{\nu}'}$ are the cubature points and their corresponding weights, 
		with $N_{\vect{\nu}'}$ being the number of cubature points. $N_{\vect{\nu}'}$ is determined by the dimension of $\vect{\nu}'$ 
		and the degree $r$ of the cubature\footnote{A cubature rule has degree $r$ if it is exact for a (multivariate) polynomial of degree $r$.}. 
		The cubature points and their weights are determined by the \gls{pdf} of $\vect{\nu}'$ and $r$. 
		Then, \eqref{eq:original problem} becomes
		\begin{IEEEeqnarray}{rCl}
			\min_{\vect{q}} \sum\nolimits_{j=1}^{N_{\vect{\nu}'}} p_j \text{SPEB}(\vect{q}, \vect{\nu}_j')\;\; \text{s.t. } & \vect{q}\succcurlyeq \vect{0},\; \vect{1}\trans\vect{q}\leq 1.
			\label{eq:full discretization problem}
		\end{IEEEeqnarray}
		In a similar fashion to \cite{KSW+19}, 
		using the epigraph form of \eqref{eq:full discretization problem}, 
		we can show that it is equivalent to the following \gls{SDP}:
		\begin{IEEEeqnarray}{rCl}
			\min_{\vect{q}, \vect{B}_1, \ldots, \vect{B}_{N_{\vect{\nu}'}}} \sum\nolimits_{j=1}^{N_{\vect{\nu}'}} p_j \trace(\vect{B}_j)\quad
			\text{s.t. }&&
			\begin{bmatrix}
				\vect{B}_j & \vect{E}\trans\\
				\vect{E} & \vect{J}(\vect{q},\vect{\nu}_j')\\
			\end{bmatrix}\succeq \vect{0}, \; j=1,\ldots, N_{\vect{\nu}'}\nonumber\\
			&&\vect{q}\succcurlyeq \vect{0},\; \vect{1}\trans\vect{q}\leq 1,
			\label{eq:full discretization problem SDP}
			\IEEEeqnarraynumspace
		\end{IEEEeqnarray}
		where $\vect{B}_j\in \mathbb{R}^{2\times 2},\;j=1,\ldots,N_{\vect{\nu}}$ are auxiliary variables of the \gls{SDP} and $\succeq$ denotes positive semidefiniteness. The positivity requirement on the cubature weights is imposed to ensure convexity of the objective in \eqref{eq:full discretization problem SDP}.
		
		The optimal vector q obtained with~\eqref{eq:full discretization problem SDP} may indicate that a very low power should be allocated in the direction of the LOS path, which may lead to a missed detection of the LOS path at the Rx. This can be avoided by ensuring that the excitation on directions around the LOS path is at least a fraction $q_{\text{th}}$ of the excitation in any other direction. 
		To this end, for a given confidence level $\kappa$, we define $\theta_{\text{T}, l, \min}^{(\kappa)}$ and $\theta_{\text{T}, l, \max}^{(\kappa)}$ as the minimum and maximum \glspl{AOD} corresponding to the \gls{2D} \gls{Rx} locations ($l = 0$) or scatterer/reflector locations ($l = 1,\ldots, L - 1$) in the $\kappa$-confidence ellipse of the respective marginal. With a uniform grid of $N_{\theta}$ possible \glspl{AOD} $\theta_{\text{T}, l, m}$ within the interval $[\theta_{\text{T}, l, \min}^{(\kappa)}, \theta_{\text{T}, l, \max}^{(\kappa)}]$ 
		\begin{IEEEeqnarray}{rCl}
			\theta_{\text{T},l,m}^{(\kappa)} &=& \theta_{\text{T}, l, \min}^{(\kappa)} + \frac{m-1}{N_{\theta} - 1}\theta_{\text{T}, l, \max}^{(\kappa)},\; m = 1, \ldots, N_{\theta}, \label{eq:angles in confidence interval}
		\end{IEEEeqnarray}
		we define the excitation matrix $\vect{A}_l\in\mathbb{R}^{N_{\theta}\times M_{\text{T}}}$ for the $l$-th path as
		\begin{IEEEeqnarray}{rCl}
			[\vect{A}_l]_{m, k} &=& |\vect{a}_{\text{T}}\trans(\theta_{\text{T}, l, m}^{(\kappa)})\vect{f}_k|^2.
		\end{IEEEeqnarray} 
		Finally, the excitation vector for the possible \glspl{AOD} of the $l$-th path is $\vect{A}_l \vect{q}$. 
		Finally, the vector with the excitation of the possible \glspl{AOD} associated with the $l$-th path is $\vect{A}_l \vect{q}$.
		We augment \eqref{eq:full discretization problem SDP} with the following linear constraints:
		\begin{IEEEeqnarray}{rCl}
			\vect{A}_0\vect{q} \succcurlyeq q_{\text{th}} \|\vect{A}\vect{q}\|_{\infty} \vect{1}_{N_{\theta}},
			\label{eq:power ratio constraint}
		\end{IEEEeqnarray}
		where $\vect{A} = [\vect{A}_0\trans,\ldots, \vect{A}_{L - 1}\trans]\trans$. We note that the constraints~\eqref{eq:power ratio constraint} can be equivalently expressed as
		\begin{IEEEeqnarray}{rCl}
			\vect{A}_0\vect{q} \succcurlyeq q_{\text{th}} e_{\max} \vect{1}_{N_{\theta}}, \quad
			\vect{A}\vect{q} \preccurlyeq e_{\max} \vect{1}_{L N_{\theta}},
		\end{IEEEeqnarray}
		with $e_{\max}$ being an auxiliary optimization variable.
		
		The main challenge with the approach described above is that $p_{\vect{\nu}}$ is a multidimensional \gls{pdf}. 
		The number of auxiliary matrices $\vect{B}_j$ and corresponding \gls{PSD} constraints in \eqref{eq:full discretization problem SDP} is equal to the number of cubature points. 
		For known cubature rules~\cite{Coo03}, the number of points is lower bounded by $(3L + 1)^{(r - 1)/2}$, which could result in very high complexity for our optimization task, as the integrand is highly non-linear and a rule with $r\geq 5$ is required for an accurate approximation.
		
		\subsection{Low-complexity sub-optimal solution}
		\subsubsection{Dimensionality reduction}
		A way to circumvent the dimensionality challenge is to use a surrogate function 
		which involves the expectation over a smaller set of parameters. 
		To this end, we first note that
		$\vect{e}_i\trans\vect{J}^{-1}\vect{e}_i,\;i = 1,2,$ 
		is a convex function of $\vect{J}$ 
		and so is the \gls{SPEB} as a sum of convex functions. 
		Splitting $\vect{\nu}'$ into any couple of vectors $\vect{\nu}_1$ and $\vect{\nu}_2$, we can write
		\begin{IEEEeqnarray}{rCl}
			\mathbb{E}_{\vect{\nu}}[\text{SPEB}(\vect{q},\vect{\nu})] &=&
			\mathbb{E}_{\vect{\nu}}\big[\trace(\vect{E}\trans\vect{J}^{-1}(\vect{q}, \vect{\nu}')\vect{E})\big]= \mathbb{E}_{\vect{\nu}_1}\big[ \mathbb{E}_{\vect{\nu}_2|\vect{\nu}_1}\big[ \trace(\vect{E}\trans\vect{J}^{-1}(\vect{q}, \vect{\nu}_1, \vect{\nu}_2)\vect{E}) \big] \big] \nonumber\\
			&\overset{(a)}{\geq}& \mathbb{E}_{\vect{\nu}_1}\big[ \trace(\vect{E}\trans (\mathbb{E}_{\vect{\nu}_2|\vect{\nu}_1}[\vect{J}(\vect{q}, \vect{\nu}_1, \vect{\nu}_2)])^{-1} \vect{E} ) \big]
			\label{eq:lower bound expected SPEB}
			\IEEEeqnarraynumspace
		\end{IEEEeqnarray} 
		where (a) follows from Jensen's inequality.
		We choose $\vect{\nu}_1 = [\vect{p}_{\text{R}}\trans, \vect{p}_{\text{s}, 1}\trans, \ldots,\vect{p}_{\text{s}, L-1}\trans]\trans$, 
		as the position parameters are the ones determining the \glspl{AOD}, which in turn determine which beams are relevant or not.
		One could optimize the lower bound on the \gls{ESPEB} in \eqref{eq:lower bound expected SPEB}, 
		as described in \eqref{eq:original problem}-\eqref{eq:full discretization problem SDP}, 
		but the number of cubature points $N_{\vect{\nu}'}$ is still lower bounded by $(2L)^{(r - 1)/2}$.
		
		\subsubsection{Power allocation as a weighted sum of per-path power allocation vectors}
		Our aim is to reduce the complexity of the optimization problem in hand. 
		We accomplish this by taking the following heuristic approach: 
		we compute a power allocation vector $\vect{q}_l,\; l=0,\ldots,L - 1$,
		considering the uncertainty regarding each path separately 
		and then weight the resulting power allocation vectors 
		in order to minimize a lower bound on the \gls{ESPEB}.
		
		More specifically, for the power allocation vector $\vect{q}_0$ we consider only the \gls{LOS} path 
		and neglect the NLOS paths and solve
		\begin{IEEEeqnarray}{rCl}
			\vect{q}_0 \hspace*{-0.05cm}=\hspace*{-0.05cm} \argmin_{\vect{q}}\;\hspace*{-0.04cm} \mathbb{E}_{\vect{p}_{\text{R}}}\hspace*{-0.03cm} \big[\hspace*{-0.05cm} \trace(\vect{E}\trans (\mathbb{E}_{|h_0|,\alpha_{\text{R}} |\vect{p}_{\text{R}}}[\hspace*{-0.02cm}\vect{J}_{\vect{\nu}_{\text{LOS}}}(\vect{q}, \vect{p}_{\text{R}}, \alpha_{\text{R}}, |h_0|)])^{-1}\hspace*{-0.03cm} \vect{E} ) \big]\quad
			&& \text{s.t. } \vect{A}_0\vect{q} \succcurlyeq q_{\text{th}, \text{LOS}} \|\vect{A}_0\vect{q}\|_{\infty} \vect{1}_{N_{\theta}},\nonumber\\
			&& \quad\;\; \vect{q}\succcurlyeq \vect{0},\; \vect{1}\trans\vect{q}\leq 1,
			\label{eq:suboptimal LOS}
		\end{IEEEeqnarray}
		where $\vect{J}_{\vect{\nu}_{\text{LOS}}}$ represents the FIM for the parameter vector 
		$\vect{\nu}_{\text{LOS}} = [\vect{p}_{\text{R}}\trans,\alpha_{\text{R}}, \epsilon_{\text{clk}}, \vect{h}_0\trans]\trans$. 
		Similarly to \eqref{eq:power ratio constraint}, the first constraint in \eqref{eq:suboptimal LOS} limits the ratio of power spent among possible \gls{LOS} directions, with $q_{\text{th},\text{LOS}}$ being the corresponding minimum ratio. 
		For the gain of the LOS path it is natural that $p(\vect{h}_0|\vect{p}_{\text{R}}) = p(\vect{h}_0|d_0)$, 
		with $d_0 =\|\vect{p}_{\text{R}}\|_2$, 
		i.e. the distribution of the gain depends only on the \gls{Tx}-\gls{Rx} distance, and 
		the integration over the radial component $d_0$ 
		and the angular component $\theta_{\text{T}, 0}$ of $\vect{p}_{\text{R}}$ can be carried out separately. 
		Then, as shown in the Appendix, 
		we can reformulate \eqref{eq:suboptimal LOS} as \pgls{SDP} using a \gls{1D} quadrature rule for the approximation of the expectation integral over $\theta_{\text{T}, 0}$.
		
		For the power allocation vector $\vect{q}_l$ we consider only the $l$-th \gls{NLOS} path 
		and assume that the \gls{Rx} position and orientation are known 
		and equal to their mean values $\bar{\vect{p}}_{\text{R}}$ and $\bar{\alpha}_{\text{R}}$. 
		This is basically a bistatic radar setup, 
		where the goal is the estimation of the point of incidence.
		Therefore, we obtain $\vect{q}_l$ by solving
		\begin{IEEEeqnarray}{rCl}
			\vect{q}_l = \argmin_{\vect{q}}\;&& \mathbb{E}_{\vect{p}_{\text{s},l}} \big[ \trace(\vect{E}\trans (\mathbb{E}_{|h_l||\vect{p}_{\text{s},l}}[\vect{J}_{\text{NLOS},l}(\vect{q}, \vect{p}_{\text{s},l}, |h_l|)])^{-1} \vect{E} ) \big]\quad \text{s.t. } \vect{q}\succcurlyeq \vect{0},\; \vect{1}\trans\vect{q}\leq 1,
			\label{eq:suboptimal NLOS}
		\end{IEEEeqnarray}
		where $\vect{J}_{\text{NLOS},l}$ represent the FIM for the parameter vector 
		$\vect{\nu}_{\text{NLOS},l} = [\vect{p}_{\text{s},l}\trans, \epsilon_{\text{clk}}, \vect{h}_l\trans]\trans$. 
		Problem~\eqref{eq:suboptimal NLOS} can be solved employing a \gls{2D} cubature on $\vect{p}_{\text{s},l}$.
		
		Finally, we compute the optimal weights $\vect{w}\in\mathbb{R}^L$ of $\vect{q}_l,l=0,\ldots,L-1$, 
		by minimizing an approximate lower bound on the \gls{ESPEB}, 
		obtained similarly to \eqref{eq:lower bound expected SPEB}:
		\begin{IEEEeqnarray}{rCl}
			\vect{w} = \argmin_{\vect{w}'}\; \mathbb{E}_{\vect{p}_{\text{R}}}[\trace(\vect{E}\trans \vect{J}^{-1}(\vect{Q}\vect{w}', \bar{\vect{\nu}}) \vect{E} )]\quad
			&& \text{s.t. } \vect{A}_0\vect{Q}\vect{w}' \succcurlyeq q_{\text{th}} \|\vect{A}\vect{Q}\vect{w}'\|_{\infty} \vect{1}_{N_{\theta}}\nonumber\\
			&& \quad \;\;\vect{Q}\vect{w}'\succcurlyeq \vect{0},\; \vect{1}\trans\vect{Q}\vect{w}'\leq 1,
			\label{eq:suboptimal all paths}
		\end{IEEEeqnarray}
		where, in order to further reduce the computational load, 
		we have replaced $\mathbb{E}_{\vect{\nu}|\vect{p}_{\text{R}}}[\vect{J}(\vect{Q}\vect{w}', \vect{\nu})]$ 
		with its approximation $\vect{J}(\vect{Q}\vect{w}', \bar{\vect{\nu}})$, 
		with $\bar{\vect{\nu}} = \mathbb{E}_{\vect{\nu}|\vect{p}_{\text{R}}}[\vect{\nu}]$ and $\vect{Q} = [\vect{q}_0,\ldots,\vect{q}_{L - 1}]$.
		Finally, the beam power allocation vector is $\vect{q} = \vect{Q}\vect{w}$. 
		

	\section{Channel and Position Estimation}
		\label{sec:position estimation}
		
		In this section we present a novel algorithm 
		for \gls{Rx} position, orientation and clock offset estimation. 
		In the first step of the algorithm a gridless parameter estimation algorithm
		based on \cite{BSR17} is employed to recover the number paths 
		and their respective \glspl{TOA}, \glspl{AOD} and \glspl{AOA}.
		In the second step, the recovered channel parameters are mapped to the position parameter vector $\vect{\nu}$.
		\subsection{Channel parameter estimation}
		
		For our positioning purposes, we are not merely interested in denoising $\vect{Y}$, 
		but we would like to recover the number of paths, 
		along with their respective gains, \glspl{TOA}, \glspl{AOD} and \glspl{AOA}.
		Hence, we aim to solve the following optimization problem:
		\begin{IEEEeqnarray}{rCl}
			\min_{L', \{\tau_l, \theta_{\text{T},l}, \theta_{\text{R},l}, h_l\}_{l=0}^{L'-1}} &&\Lambda(\vect{R}) + \chi \|\vect{h}\|_1  \label{eq:optimization problem final form}
		\end{IEEEeqnarray}
		where
		\begin{IEEEeqnarray}{rCl}
			\Lambda(\vect{R}) = \frac{1}{2}\|\vect{R}\|_{\text{F}}^2\label{eq:loss function definition}
		\end{IEEEeqnarray} 
		is the loss function, 
		\begin{IEEEeqnarray}{rCl}
			\vect{R} = \vect{Y} - \sum\nolimits_{l=0}^{L'-1} h_l \vect{C}(\tau_l, \theta_{\text{T},l} \theta_{\text{R},l})\label{eq:residual definition}		
		\end{IEEEeqnarray} 
		is the residual, $\chi$ is a regularization parameter and $\vect{h} = [h_0, \ldots, h_{L' - 1}]\trans$. 
		The penalty term $\|\vect{h}\|_1$ is included to make the channel representation more parsimonious; 
		otherwise the number of detected paths could grow arbitrarily so as to minimize the objective. 
		As usual in sparse recovery setups, 
		instead of a non-convex L0 norm penalty term, 
		we use the L1 norm.
		We solve problem~\eqref{eq:optimization problem final form} 
		using the algorithmic framework of \cite{BSR17}, 
		termed as \gls{ADCGM}, which is described in Alg.~\ref{alg:ADCGM}. 
		We note that, for notational brevity, 
		in \eqref{eq:optimization problem final form}-\eqref{eq:residual definition} and in the following,
		we write $\vect{R}$ instead of $\vect{R}(L', \{\tau_l, \theta_{\text{T}, l}, \theta_{\text{R}, l}\}_{l=0}^{L'-1})$. Also, 
		the residual at iteration $i$ is denoted as $\vect{R}_i$ 
		and the \glspl{TOA} of the detected paths are stacked in the vector
		$\vect{\tau}^{(i)} = [\tau_0^{(i)}, \ldots, \tau_{L^{(i)} - 1}^{(i)}] \in \mathbb{R}^{L^{(i)}}$, 
		where $L^{(i)}$ is the number of detected paths at iteration $i$. 
		The parameter vectors $\vect{\theta_{\text{T}}}^{(i)}$ and $\vect{\theta_{\text{R}}}^{(i)}$ are defined accordingly.	
		The maximum number of iterations is $L_{\max}$ and at each iteration a new path can be detected (Step 2) or previously detected paths can be dropped (Step 4(b)).
		\begin{algorithm}
			\caption{Channel parameter estimation with \gls{ADCGM}}
			\label{alg:ADCGM}
			\begin{algorithmic}
				\Input{$\{\vect{X}_b\}_{b=1}^{N_{\text{B}}},\; \vect{Y},\; \sigma_{\eta}^2,\; P_{\text{fa}}$}
				\Initialize{$\vect{\tau}^{(0)}, \vect{\theta}_{\text{T}}^{(0)}, \vect{\theta}_{\text{R}}^{(0)}, \vect{h}^{(0)} = [\;]$}
				\Do
				\State 1. Compute residual $\vect{R}_i$ 
				\State 2. Detect next potential path:
				\begin{IEEEeqnarray}{rCl}
					\tau^{(i)}, \theta_{\text{T}}^{(i)}, \theta_{\text{R}}^{(i)} = \argmax_{(\tau, \theta_{\text{T}}, \theta_{\text{R}})\in \mathcal{G}} \big|\trace(\vect{R}_i\herm \vect{C}(\tau, \theta_{\text{T}}, \theta_{\text{R}}))\big|\label{eq:next source problem}
				\end{IEEEeqnarray}
				\State 3. 
				Update support: 
				$\vect{\tau}^{(i + 1)} = [(\vect{\tau}^{(i)})\trans, \tau^{(i)}]$, \newline \hspace*{0.85cm} $\vect{\theta}_{\text{T}}^{(i + 1)} = [(\vect{\theta}_{\text{T}}^{(i)})\trans, \theta_{\text{T}}^{(i)}],\; \vect{\theta}_{\text{R}}^{(i + 1)} = [(\vect{\theta}_{\text{R}}^{(i)})\trans, \theta_{\text{R}}^{(i)}]$
				\State 4. Coordinate descent on non-convex objective:
				\For{$i=1$ to $N_{\text{cd}}$}
				\State (a) Compute gains: 
				\begin{IEEEeqnarray}{rCl}
					\vect{h}^{(i + 1)} = \argmin\nolimits_{\vect{h}} \Lambda(\vect{R}) + \chi \|\vect{h}\|_1 \label{eq:sparse gain update}
				\end{IEEEeqnarray}
				\State (b) Prune support: 
				\begin{IEEEeqnarray*}{rCl}
					\{\vect{\tau}, \vect{\theta}_{\text{T}}, \vect{\theta}_{\text{R}}, \vect{h}\}^{(i + 1)} = \mathrm{prune}(\{\vect{\tau}, \vect{\theta}_{\text{T}}, \vect{\theta}_{\text{R}}, \vect{h}\}^{(i + 1)})
				\end{IEEEeqnarray*}
				\State (c) Locally improve support: 
				\begin{IEEEeqnarray*}{rCl}
					\{\vect{\tau}, \vect{\theta}_{\text{T}}, \vect{\theta}_{\text{R}}\}^{(i + 1)}=\mathrm{local\_descent}(\{\vect{\tau}, \vect{\theta}_{\text{T}}, \vect{\theta}_{\text{R}}, \vect{h}\}^{(i + 1)})	
				\end{IEEEeqnarray*} 
				\EndFor
				\State $i = i + 1$
				\doWhile{$k < L_{\max}$ and $\big|\trace(\vect{R}_i\herm \vect{C}(\tau^{(i)}, \theta_{\text{T}}^{(i)}, \theta_{\text{R}}^{(i)})\big| > \zeta_1$}
			\end{algorithmic}
		\end{algorithm}
		In the following, we describe steps 2 and 4 in detail.
		
		\subsubsection{Detection of a new potential path (Step 2)}
		In order to get the next potential path we have to solve~\eqref{eq:next source problem}, 
		which is 
		non-convex and can be solved by discretizing the \gls{3D} parameter space $[0, T_{\text{CP}}]\times [-\pi, \pi) \times [-\pi, \pi)$ to get an $N_{\tau}\times N_{\theta_{\text{T}}}\times N_{\theta_{\text{R}}}$-dimensional grid $\mathcal{G}$.
		
		After computing the new potential source, 
		we compare the correspoding objective with a predefined threshold $\zeta_1>0$, 
		which is a function of the noise variance $\sigma_{\eta}^2$, the reference signal $\vect{X}$ and the desired false alarm probability $P_{\text{fa}}$.
		
		\subsubsection{Coordinate descent (Step 4)}
		In this algorithmic step we iteratively perform 3 sub-steps for a fixed number of $N_{\text{cd}}$ iterations:
		\begin{enumerate}[label = (\alph*)]
			\item We update the gains solving~\eqref{eq:sparse gain update}, keeping the other path parameters fixed. 
			The regularization parameter $\chi$ determines the accuracy-sparsity trade-off.  
			\item We prune the paths whose gain is effectively zero: 
			the $l$-th path is pruned if $|h_l|^2/\zeta_2 <\max_{l=0,\ldots,L^{(i)} - 1} |h_l|^2$, where $0<\zeta_2\ll 1$.
			\item For the local descent step we perform truncated Newton steps for each path and each parameter sequentially:
			\begin{IEEEeqnarray}{rCl}
				\tau_l^{(i + 1)} &\leftarrow& \tau_l^{(i + 1)} - \mathrm{sgn}(\partial\Lambda/\partial\tau_l^{(i + 1)}) s_{\tau, l}^{(i + 1)}\\ 
				\theta_{\text{T}, l}^{(i + 1)} &\leftarrow& \theta_{\text{T},l}^{(i + 1)} - \mathrm{sgn}(\partial\Lambda/\partial\theta_{\text{T},l}^{(i + 1)}) s_{\theta_{\text{T}}, l}^{(i + 1)}\\
				\theta_{\text{R}, l}^{(i + 1)} &\leftarrow& \theta_{\text{R},l}^{(i + 1)} - \mathrm{sgn}(\partial\Lambda/\partial\theta_{\text{R},l}^{(i + 1)}) s_{\theta_{\text{R}}, l}^{(i + 1)} 
			\end{IEEEeqnarray}
			where 
			\begin{IEEEeqnarray}{rCl}
				s_{\tau, l}^{(i + 1)} &=& \min \bigg(\Big|\Big(\partial^2\Lambda/(\partial\tau_l^{(i + 1)})^2\Big)^{-1} \partial\Lambda/\partial\tau_l^{(i + 1)}\Big|, \;\frac{N_{\text{CP}} T_{\text{s}}}{2(N_{\tau} - 1)}  \bigg) \nonumber\\
				s_{\theta_{\text{T}}, l}^{(i + 1)} &=& \min \bigg(\Big|\Big(\partial^2\Lambda/(\partial\theta_{\text{T}, l}^{(i + 1)})^2\Big)^{-1} \partial\Lambda/\partial\theta_{\text{T}, l}^{(i + 1)}\Big|, \; \frac{\pi}{N_{\theta_{\text{T}} - 1}}  \bigg) \nonumber\\
				s_{\theta_{\text{R}}, l}^{(i + 1)} &=& \min \bigg(\Big|\Big(\partial^2\Lambda/(\partial\theta_{\text{R}, l}^{(i + 1)})^2\Big)^{-1} \partial\Lambda/\partial\theta_{\text{R}, l}^{(i + 1)}\Big|, \; \frac{\pi}{N_{\theta_{\text{R}} - 1}}  \bigg) \nonumber
			\end{IEEEeqnarray}
			are the step sizes, with $T_{\text{s}} = N\Delta f$.
			We note that we limit the maximum step size for each of the parameters to be equal to half of the corresponding grid bin size, in order to avoid convergence problems near inflection points of the loss function.
			
		\end{enumerate}
		
		\subsection{Mapping to position parameters}
		Having an estimate $\hat{\tilde{\vect{\nu}}}$ of the channel parameter vector $\tilde{\vect{\nu}}$ defined as
		\begin{IEEEeqnarray}{rCl}
			\tilde{\vect{\nu}} &=& [\tau_0, \theta_{\text{T}, 0}, \theta_{\text{R}, 0},\ldots, \tau_{\hat{L} - 1}, \theta_{\text{T},\hat{L} - 1}, \theta_{\text{R},\hat{L} - 1}]\trans,
		\end{IEEEeqnarray}
		where $\hat{L}$ is the estimated number of paths, and choosing the strongest path as the \gls{LOS} path,
		we estimate the position parameter vector $\vect{\nu}$ employing the \gls{EXIP} as in~\cite{SGD+18}, 
		with a slight modification to include the prior information on the clock offset. 
		To this end, we intend to solve
		\begin{IEEEeqnarray}{rCl}
			\argmin_{\vect{\nu}} \;(\hat{\tilde{\vect{\nu}}} - f(\vect{\nu}))\trans \vect{J}_{\hat{\tilde{\vect{\nu}}}}(\hat{\tilde{\vect{\nu}}} - f(\vect{\nu})) + (\epsilon_{\text{clk}}/\sigma_{\text{clk}})^2, \label{eq:optimization channel to position}
		\end{IEEEeqnarray} 
		where $\vect{J}_{\hat{\tilde{\vect{\nu}}}}$ is the channel parameter \gls{FIM}
		and $f:\mathbb{R}^{2\hat{L} + 2} \rightarrow \mathbb{R}^{3\hat{L}}$ is the mapping from position to channel parameters, 
		determined by~\eqref{eq:pos2ch tau},~\eqref{eq:pos2ch thT}-\eqref{eq:pos2ch thR}. 
		
		We note that false alarms, that is falsely detected paths, can have severe impact on position estimation.
		Therefore, we apply the following two criteria to filter them out:
		\begin{itemize}
			\item A single-bounce \gls{NLOS} path and a LOS path always form a triangle, as can be seen in Fig.~\ref{fig:geometric_model}. Therefore, for the formation of a triangle to be possible, a single-bounce \gls{NLOS} path must satisfy
			\begin{IEEEeqnarray}{rCl}
				\Delta\theta_{\text{T}, l} \cdot \Delta\theta_{\text{R},l} < 0,\; l = 1, \ldots, \hat{L} - 1, \label{eq:condition for single-bounce paths}
			\end{IEEEeqnarray}
			where $\Delta\theta_{\text{T}, l} = \theta_{\text{T}, l} - \theta_{\text{T}, 0}$ and $\Delta\theta_{\text{R},l} = \theta_{\text{R},l} - \theta_{\text{R},0}$, with $\Delta\theta_{\text{T}, l}$ and $\Delta\theta_{\text{R}, l} \in [-\pi, \pi)$. 
			Therefore if the $l$-th path, $l = 1, \ldots, \hat{L} - 1$, does not satisfy~\eqref{eq:condition for single-bounce paths}, it is dropped.
			\item Combined with the LOS path, each NLOS path can provide an estimate of $\epsilon_{\text{clk}}$:
			\begin{IEEEeqnarray}{rCl}
				\epsilon_{\text{clk}, l} \hspace*{-0.05cm}=\hspace*{-0.05cm} \frac{\tau_l \sin(\Delta\theta_{\text{R},l} - \Delta\theta_{\text{T},l}) \hspace*{-0.02cm}-\hspace*{-0.02cm} \tau_0(\sin(\Delta\theta_{\text{R},l})\hspace*{-0.02cm} - \hspace*{-0.02cm} \sin(\Delta\theta_{\text{T},l}))}{\sin(\Delta\theta_{\text{R},l} - \Delta\theta_{\text{T},l}) \hspace*{-0.02cm}-\hspace*{-0.02cm} (\sin(\Delta\theta_{\text{R},l})\hspace*{-0.02cm} - \hspace*{-0.02cm} \sin(\Delta\theta_{\text{T},l}))},
				\IEEEeqnarraynumspace
			\end{IEEEeqnarray} 
			With $\zeta_{3,a} > 0$ and $\zeta_{3,b} > 0$ being predefined probability thresholds for $\epsilon_{\text{clk}}$ values, if $p(\epsilon_{\text{clk},l}) < \zeta_{3,a}$ or $p(\epsilon_{\text{clk},l}) < \zeta_{3,b}p_{\text{clk},\max}$, the path is filtered out, with $p_{\text{clk},\max} = \max_{l=1,\ldots,\hat{L} - 1}p(\epsilon_{\text{clk},l})$.
		\end{itemize}
		
		Replacing $\hat{\tilde{\vect{\nu}}}$ with $\hat{\tilde{\vect{\nu}}}'$, 
		which contains only the remaining paths, 
		we solve~\eqref{eq:optimization channel to position} with the Levenberg-Marquardt algorthm~\cite{Lev44, Mar63}.
		For the initial point $\vect{\nu}^{(0)}$ we compute
		\begin{IEEEeqnarray}{rCl}
			\epsilon_{\text{clk}}^{(0)} &=& \frac{\sum_l |h_l|^2 \epsilon_{\text{clk}, l}}{\sum_l |h_l|^2}\\
			\vect{p}_{\text{R}}^{(0)} &=& c(\tau_0 - \epsilon_{\text{clk}}^{(0)}) \vect{u}(\theta_{\text{T}, 0})\\
			\alpha_{\text{R}}^{(0)} &=& \theta_{\text{T}, 0} + \pi - \theta_{\text{R}, 0}\\
			\vect{p}_{\text{s},l}^{(0)} &=& \frac{\tan(\theta_{\text{R},l} + \alpha_{\text{R}}^{(0)})p_{\text{R}, x}^{(0)} - p_{\text{R}, y}^{(0)}}{\tan(\theta_{\text{R},l} + \alpha_{\text{R}}^{(0)})\cos\theta_{\text{T}, l} - \sin\theta_{\text{T}, l}} \vect{u}(\theta_{\text{T}, l}), l = 1,\ldots,\hat{L}',
			\IEEEeqnarraynumspace
		\end{IEEEeqnarray}
		where $\hat{L}'$ is the number of remaining estimated paths.

	\section{Numerical Results}
		\label{sec:numerical results}
		
		\subsection{Simulation setup}
		\subsubsection{Geometric setup and prior information at the \gls{Tx}}
\begin{figure}
			\centering
			\includegraphics[scale=0.9]{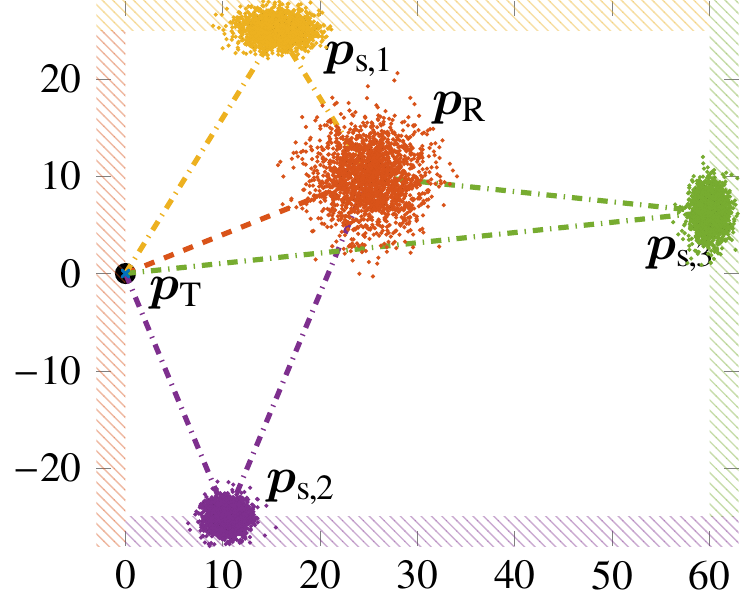}
			\caption{Prior knowledge at the Tx for simulation results.}
			\label{fig:indoor_scenario_and_prior}
		\end{figure}

		For the evaluation of the power allocation and the position estimation algorithms we consider the setup shown in Fig. \ref{fig:indoor_scenario_and_prior}. 
		The \gls{Tx} is equipped with a \gls{ULA} with $N_{\text{T}} = 32$ antennas. In order to be able to discriminate all possible \glspl{AOA}, the \gls{Rx} has a \gls{UCA} with $N_{\text{R}} = 16$ antennas. With the \gls{Rx} being equipped with a \gls{UCA}, 
		the \gls{SPEB} is independent of the orientation $\alpha_{\text{R}}$.
		
		We consider \gls{NLOS} paths resulting from single-bounce reflections. The phases of the complex path gains are uniformly distributed over $[-\pi, \pi)$ and their magnitudes are given by 
		\begin{IEEEeqnarray}{rCl}
			|h_l| = \begin{cases}
				c/(4\pi f_{\text{c}} \|\vect{p}_{\text{R}}\|_2), & l=0,\\
				\sqrt{\rho_l} c/(4\pi f_{\text{c}}(\|\vect{p}_{\text{s}, l}\|_2 + \|\vect{p}_{\text{R}} - \vect{p}_{\text{s}, l}\|_2)), & l\neq 0,
			\end{cases}
		\end{IEEEeqnarray}
		where $\rho_l$ is the reflection coefficient and $\lambda_{\text{c}} = c/f_{\text{c}}$.
		The prior knowledge at the \gls{Tx} is described by $\mathcal{N}(\vect{\mu}, \vect{C})$, where
		\begin{IEEEeqnarray}{rCl}
			\vect{\mu} &=& [\bar{\vect{p}}_{\text{R}}\trans, \bar{\vect{p}}_{\text{s}, 1}\trans, \bar{\rho},  \bar{\vect{p}}_{\text{s}, 2}\trans, \bar{\rho}, \bar{\vect{p}}_{\text{s}, 3}\trans, \bar{\rho}]\trans\in\mathbb{R}^{11} \label{eq:numerical results mean}\\
			\vect{C} &=& 
			\begin{bmatrix}
				\vect{C}_{0,0}& \vect{C}_{0,1} & \vect{0} & \vect{C}_{0, 2}& \vect{0}& \vect{C}_{0, 3} & \vect{0}\\
				\vect{C}_{0,1}\trans & \vect{C}_{1, 1} & \vect{0}& \vect{0} & \vect{0} & \vect{0} & \vect{0}\\
				\vect{0} & \vect{0} & \sigma_{\rho}^2 & \vect{0} & 0 & \vect{0} & 0\\
				\vect{C}_{0,2}\trans & \vect{0} & \vect{0} & \vect{C}_{2, 2} & \vect{0} & \vect{0} & \vect{0}\\
				\vect{0} & \vect{0} & 0 & \vect{0} & \sigma_{\rho}^2 & \vect{0} & 0\\
				\vect{C}_{0,3}\trans & \vect{0} & \vect{0} & \vect{0} & \vect{0} & \vect{C}_{3, 3} & \vect{0}\\
				\vect{0} & \vect{0} & 0 & \vect{0} & 0 & \vect{0} & \sigma_{\rho}^2\\
			\end{bmatrix}\in\mathbb{R}^{11\times 11} \label{eq:numerical results covariance}
			\IEEEeqnarraynumspace
		\end{IEEEeqnarray}
		with 
		\begin{IEEEeqnarray}{rCl}
			\bar{\vect{p}}_{\text{R}} &=& \begin{bmatrix}
				25\\
				10
			\end{bmatrix}\si{\meter},\; \vect{C}_{0, 0}= 4/\sqrt{2}\vect{I}_{2}\si{\square\meter}\nonumber\\
			\bar{\vect{p}}_{\text{s}, 1} &=& \begin{bmatrix}
				15.63\\
				25
			\end{bmatrix}\si{\meter},\; \vect{C}_{1, 1}= \begin{bmatrix}
				3.48 & 0\\
				0 & 1\\
			\end{bmatrix}\si{\square\meter}, \; \vect{C}_{0, 1}= \begin{bmatrix}
				4.45 & 0\\
				0 & 0\\
			\end{bmatrix}\si{\square\meter}\nonumber\\
			\bar{\vect{p}}_{\text{s}, 2} &=& \begin{bmatrix}
				10.42\\
				-25
			\end{bmatrix}\si{\meter},\; \vect{C}_{2, 2}= \begin{bmatrix}
				1.34 & 0\\
				0 & 1\\
			\end{bmatrix}\si{\square\meter}, \; \vect{C}_{0, 2}= \begin{bmatrix}
				1.64 & 0\\
				0 & 0\\
			\end{bmatrix}\si{\square\meter}\nonumber\\
			\bar{\vect{p}}_{\text{s}, 3} &=& \begin{bmatrix}
				60\\
				6.32
			\end{bmatrix}\si{\meter},\; \vect{C}_{3, 3}= \begin{bmatrix}
				1 & 0\\
				0 & 2.31\\
			\end{bmatrix}\si{\square\meter}, \; \vect{C}_{0, 3}= \begin{bmatrix}
				0 & 0\\
				0 & 3.24\\
			\end{bmatrix}\si{\square\meter}\nonumber\\
			\bar{\rho} &=& -10 \si{\decibel},\;\sigma_{\rho} = 4\si{\decibel}.\nonumber
		\end{IEEEeqnarray}
		Samples from this distribution are depicted in Fig.~\ref{fig:indoor_scenario_and_prior}.
		
		\subsubsection{System Parameters}
		\label{sec:numerical results, system parameters}
		For the waveform we set $f_{\text{c}} = \SI{38}{\giga\hertz},\; N = 64,\; N_{\text{B}} = 10, \; \mathcal{P} = \{-31,\ldots,-1,1\ldots, 31\}$ and $\Delta f (\max(\mathcal{P}) - \min(\mathcal{P})) (\approx B) = \SI{120}{\mega\hertz}$. The resources are assigned to the beams in an interleaved and staggered manner, i.e. $\mathcal{R}_k = \{(k + b + i M_{\text{T}}, b) | i\in\mathbb{Z}, b = 1,\ldots, N_{\text{B}} :k + b + i M_{\text{T}}\in \mathcal{P}\}$. The power of each beam is distributed uniformly among its resources, i.e., $\gamma_k[p,b] = 1/|\mathcal{R}_k|$. The noise variance is $\sigma_{\eta}^2 =  10^{0.1 (n_{\text{Rx}} + N_0)} N \Delta f $, where $N_0=\SI{-174}{\dBm\per\hertz}$ is the noise power spectral density per dimension and $n_{\text{Rx}} = \SI{8}{\decibel}$ is the \gls{Rx} noise figure. The standard deviation of the clock offset is $\sigma_{\text{clk}} = 2/(N \Delta f)$, so that $c \sigma_{\text{clk}} \approx \SI{4.88}{\meter}$. We use a DFT beam codebook:
		\begin{IEEEeqnarray}{rCl}
			\vect{f}_k = \big[1,\e^{-\jj \frac{2 \pi}{N_{\text{T}}}k}, \ldots, \e^{-\jj \frac{2 \pi}{N_{\text{T}}}(N_{\text{T}} - 1)k}\big], \; k = 1,\ldots, M_{\text{T}} = N_{\text{T}}.
		\end{IEEEeqnarray}
		
		\subsubsection{Benchmark for beam power allocation}
		In order to fairly evaluate our power allocation strategies, we set as benchmark the uniform power allocation to beams exciting useful directions. For a given confidence level $\kappa$ we get a grid of \glspl{AOD} for each path as in \eqref{eq:angles in confidence interval} and compute the set of useful beams as
		\begin{IEEEeqnarray}{rCl}
			\mathcal{B}_{\text{uni}}^{(\kappa)} &=& \cup_{l=0}^{L - 1}\cup_{m=0}^{N_{\theta}} \Big\{\argmax_{k=1,\ldots,N_{\text{T}}}|\vect{a}_{\text{T}}\trans(\theta_{\text{T}, l, m}^{(\kappa)})\vect{f}_k|\Big\}. 
			\label{eq:useful beams}
		\end{IEEEeqnarray}  
		The power allocation vector $\vect{q}$ is
		\begin{IEEEeqnarray}{rCl}
			q_k = \begin{cases}
				1/|\mathcal{B}_{\text{uni}}^{(\kappa)}|, & k \in \mathcal{B}_{\text{uni}}^{(\kappa)}\\
				0, & k \notin \mathcal{B}_{\text{uni}}^{(\kappa)}.
			\end{cases}
			\label{eq:useful uniform power allocation}
		\end{IEEEeqnarray}
		
		\subsection{Power allocation strategies and position estimation algorithm parameters}
		\label{sec:results - power allocation strategies}
\begin{figure}
			\centering
			\subfloat[opt. unconstr.]{\begin{adjustbox}{scale=0.51}
					\includegraphics[scale=1]{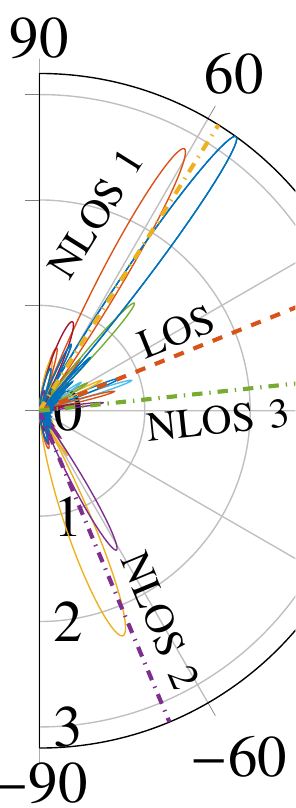}\hspace*{1.2cm}
			\end{adjustbox}}\qquad
			\subfloat[opt. constr.]{\begin{adjustbox}{scale=0.51}
					\includegraphics[scale=1]{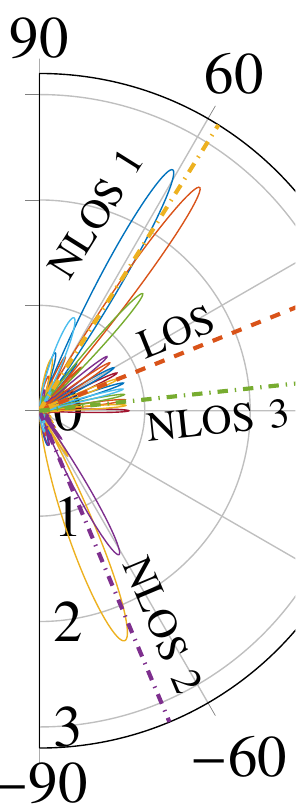}\hspace*{1cm}
			\end{adjustbox}}\qquad
			\subfloat[opt. reduced]{\begin{adjustbox}{scale=0.51}
					\includegraphics[scale=1]{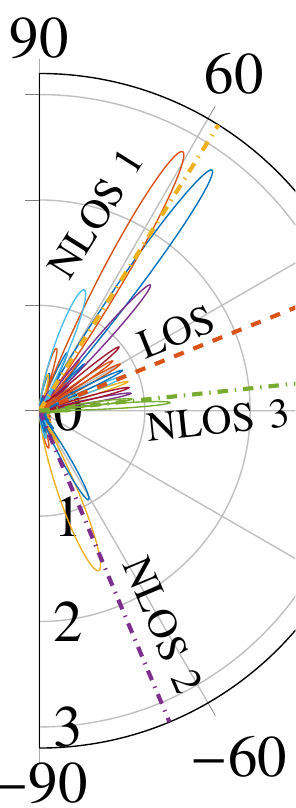}\hspace*{1cm}
			\end{adjustbox}}
			
			\subfloat[subopt.]{\begin{adjustbox}{scale=0.51}
					\includegraphics[scale=1]{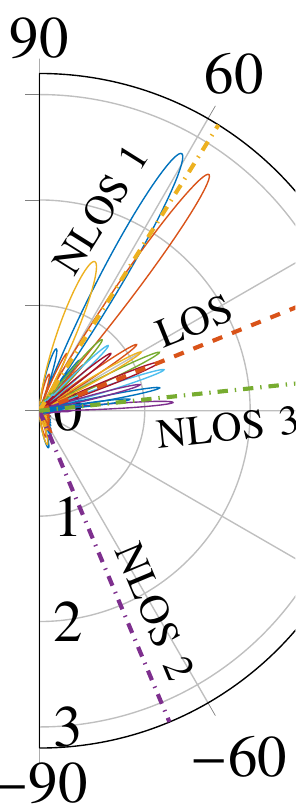}\hspace*{1.2cm}
			\end{adjustbox}}\qquad
			\subfloat[uni 0.60]{\begin{adjustbox}{scale=0.51}
					\includegraphics[scale=1]{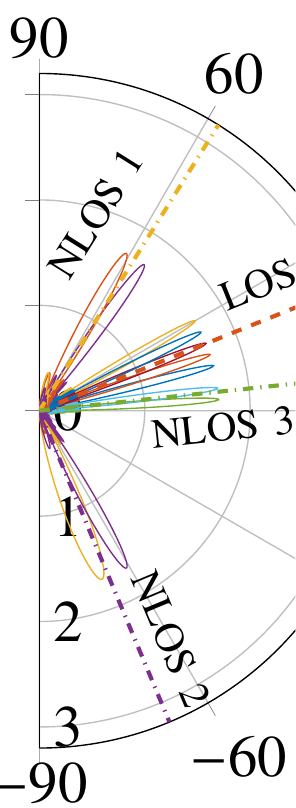}\hspace*{1cm}
			\end{adjustbox}}\qquad
			\subfloat[uni 0.90]{\begin{adjustbox}{scale=0.51}
					\includegraphics[scale=1]{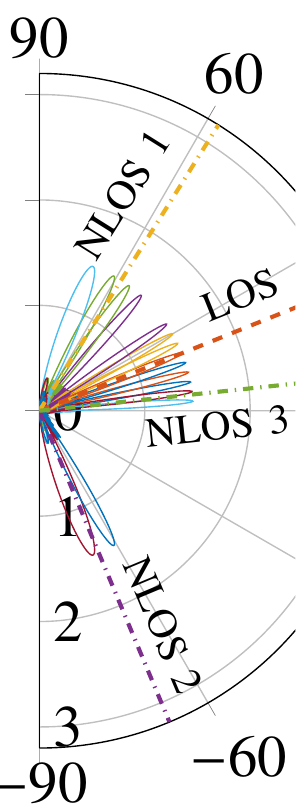}\hspace*{1cm}
			\end{adjustbox}}
			\caption{Beam patterns $|\vect{a}_{\text{T}}\trans(\theta_{\text{T}})\vect{f}_k\sqrt{q_k}|, k=1,\ldots,M_{\text{T}}$, for different power allocation strategies.}
			\label{fig:beampatterns}
		\end{figure}

		The power allocation strategies and their corresponding parametrizations that we consider for our simulation results are as follows:
		\begin{itemize}
			\item \textit{opt. unconstr.}: Solution of \eqref{eq:full discretization problem SDP}. The number of points of known cubatures of \nth{5} degree (in order to ensure a sufficiently dense sampling of the support of the distribution) with positive weights is $2^{11} + 2\cdot 11 = 4118$, which incurs prohibitive computational complexity. Instead, we draw $11^2=121$ random samples (as many as the lower bound for any cubature) from the joint $11$-dimensional distribution.
			\item\textit{opt. constr.}: Solution of \eqref{eq:full discretization problem SDP} with $121$ random samples from the joint $11$-dimensional distribution and additional constraints \eqref{eq:power ratio constraint}, with $\kappa = 0.995$, $q_{\text{th}} = \SI{-10}{\decibel}$ and $N_{\theta} = 15$.
			\item \textit{opt. reduced}: Solution of the minimization of the lower bound on \gls{ESPEB} \eqref{eq:lower bound expected SPEB} with $8^2=64$ random samples from the joint $8$-dimensional distribution and additional constraints \eqref{eq:power ratio constraint}, with $\kappa = 0.995$, $q_{\text{th}} = \SI{-10}{\decibel}$ and $N_{\theta} = 15$.
			\item \textit{subopt.}: Solution of \eqref{eq:suboptimal LOS}-\eqref{eq:suboptimal all paths}, with $9$-point cubatures for the involved \gls{2D} marginals, $\kappa = 0.995$, $q_{\text{th}, \text{LOS}} = \SI{-3}{\decibel}$, $q_{\text{th}} = \SI{-10}{\decibel}$ and $N_{\theta} = 15$.
			\item \textit{uni $\kappa$}: Uniform power allocation to useful directions, according to \eqref{eq:useful beams}-\eqref{eq:useful uniform power allocation}, with $\kappa = \{0.60,\;0.90\}$ and $N_{\theta} = 15$.	We note that choosing $\kappa=0.995$ as for the other strategies results in performance degradation; hence, results for this value are not icluded.
		\end{itemize}
		The beampatterns of the power allocation strategies for the considered prior knowledge are shown in Fig.~\ref{fig:beampatterns}.
		We observe in Figs.~\ref{fig:beampatterns}(a)-(d) that for the optimized power allocation strategies, most of the available power is spent on beams illuminating \gls{NLOS} paths. 
		When $\sigma_{\text{clk}}$ is very small (i.e. when the synchronization error is very small), having only the delay measurement of the \gls{LOS} suffices to determine the distance between the \gls{BS} and the \gls{UE}. However, as $\sigma_{\text{clk}}$ increases, neither the \gls{LOS} nor the \gls{NLOS} provide individually information about the \gls{BS}-\gls{UE} distance. In these cases, they are the differences between delays that are informative, and this implies that several paths (not only one) have to be illuminated with sufficient power because if there is a large power unbalance between rays, then the delay differences will not be estimated precisely.
		Comparing  Fig.~\ref{fig:beampatterns}(a) with Figs.~\ref{fig:beampatterns}(b)-(d), we see that when the constraints~\eqref{eq:power ratio constraint} are not applied, the power allocation to \gls{NLOS} components is more significant, with the power invested to less likely \gls{LOS} directions being very low. From Figs.~\ref{fig:beampatterns}(b) and (c), we can see that the impact of the dimensionality reduction~\eqref{eq:lower bound expected SPEB} is the reduction of the power spent on the \nth{2} \gls{NLOS} path. This is explained by the fact that the fading of the path gains is not taken into account; hence, for the mean values of the path gains, more power is spent on the paths that offer more useful position information. Also, in Fig.~\ref{fig:beampatterns}(d) we observe that our suboptimal approach allocates almost no power to the \nth{2} NLOS path, as in the last step where all paths are considered jointly, only the receiver's location uncertainty and the mean scatterers'/reflectors' locations are taken into account; for this setup, the information offered by the \nth{1} \gls{NLOS} path is more useful and therefore most of the available power is allocated for its illumination. For the uniform allocation, higher confidence values lead to activation of more beams and spreading of the avaiable power to more directions.
		
		Regarding the position estimation algorithm parameters, we set $N_{\tau} = 2 N_{\text{P}},\; N_{\theta_{\text{T}}} = 2 N_{\text{T}},\; N_{\theta_{\text{R}}} = 2 N_{\text{R}},\; P_{\text{fa}} = 0.05$, $\zeta_1$ is pre-trained for the given $P_\text{fa}$ and power allocation strategy, $\zeta_2 = -\SI{35}{\decibel}$, $N_{\text{cd}} = 3,\; L_{\max} = 10,\; \chi = \sigma_{\eta}\sqrt{2(N_{\text{T}} + N_{\text{R}}) |\mathcal{P}|N_{\text{B}}P_{\text{RE}}/N_{\text{T}}}$ (chosen according to~\cite{ZGS+15}), $\zeta_{3,a} = 10^{-4}$ and $\zeta_{3,b} = 10^{-2}$.

		\subsection{Performance vs SNR for fixed geometry}
\begin{figure}
			\centering
			\includegraphics[scale=0.95]{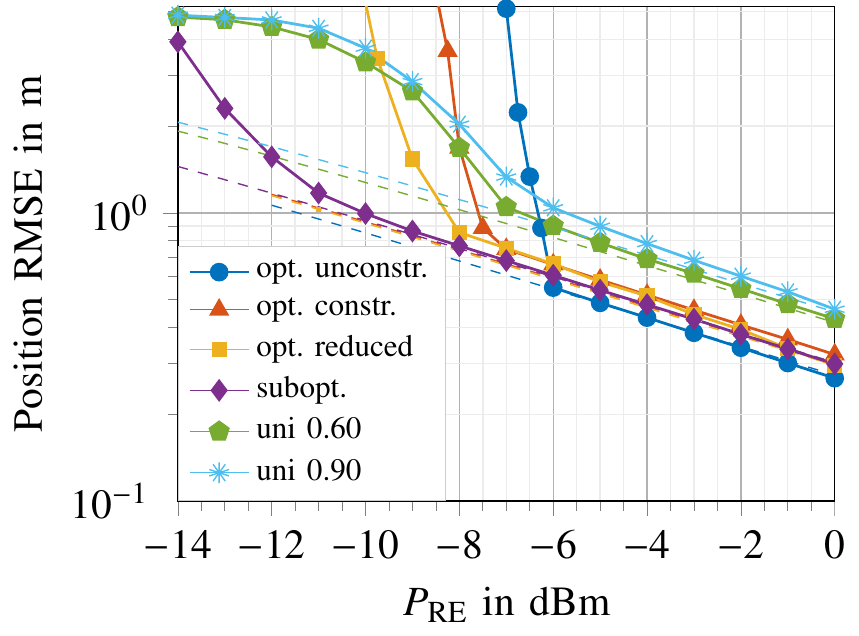}
			\caption{Position \gls{RMSE} (solid lines) and \gls{PEB} (dashed lines) vs \gls{Tx} power for different power allocation strategies.}
			\label{fig:position_RMSE_and_RCRLB_vs_tx_power}
		\end{figure}

		We fix the geometry and the reflection coefficients to their mean value $\vect{\mu}$ in~\eqref{eq:numerical results mean} to examine the performance of the position estimation algorithm as a function of the \gls{Tx} power. For the power allocation strategies described in Sec.~\ref{sec:results - power allocation strategies}, in Fig.~\ref{fig:position_RMSE_and_RCRLB_vs_tx_power}, we plot the position \gls{RMSE} $\mathbb{E}_{\vect{\eta}, \epsilon_{\text{clk}}}[\|\hat{\vect{p}}_{\text{R}} - \vect{p}_{\text{R}}\|_2^2]$ and \gls{PEB} as functions of the average power per resource element $P_{\text{RE}} = P_{\text{tot}}/(N_{\text{B}}N_{\text{P}})$, with $\hat{\vect{p}}_{\text{R}}$ being the position estimate. We note that the average \gls{Tx} power $P_{\text{T}}$ is related to $P_{\text{RE}}$ as $P_{\text{T}} = P_{\text{RE}}N_{\text{P}}/N$.
		
		We can see that the bound is attained for all power allocation strategies. 
		Regarding uniform power allocation, the distance of the \gls{RMSE} from the bound for low \gls{Tx} power is attributed to the fact that, although the LOS path is detected, the probability of detection for the \gls{NLOS} is small. 
		With only the LOS path being detected, the clock offset cannot be resolved
		and the resulting position \gls{RMSE} approaches the standard deviation of the clock offset $c\cdot\sigma_{\text{clk}}\approx \SI{4.88}{\meter}$. Among the two considered configurations ($\kappa = 0.60$ and $\kappa = 0.90$), the former has slightly better performance, as the available power is more concentrated to the true location of the \gls{Rx} and the reflectors. But, as we will see later on, this comes with a cost, when the uncertainty about the geometry is considered.
		
		The optimized allocation strategies result in similar \glspl{PEB} and offer significant improvement compared to the uniform, with a gain of $3$ to $4$ \si{\decibel} for the same localization accuracy. The lowest \gls{PEB} is attained by "opt. unconstr.", but the \gls{RMSE} converges to the PEB for larger $P_{\text{RE}}$, compared to the other strategies. The reason for this behavior is that, as can be observed in Fig.~\ref{fig:beampatterns}(a), only a small fraction of power is spent in the LOS direction and the Tx power required for the LOS path to be detected is larger. When the LOS path is missed, the first arriving NLOS path is treated as LOS by the algorithm, resulting in a large position error. 
		Due to the constraints~\eqref{eq:power ratio constraint}, the rest of the proposed strategies ("opt. constr.", "opt. reduced" and "subopt.") allocate more power to the LOS, enabling to attain the PEB at lower values of $P_{\text{RE}}$, with only a small performance penatly.
		The \gls{RMSE} of "opt. reduced" converges slightly faster to the bound compared to "opt. constr.", as slightly more power is allocated to the LOS path. The "subopt." allocation exhibits the most robust performance, as the \gls{LOS} 
		path can be detected for much lower Tx power values. 
		
		\subsection{Performance with random geometry}
\begin{figure}
			\centering
			\includegraphics[scale=0.90]{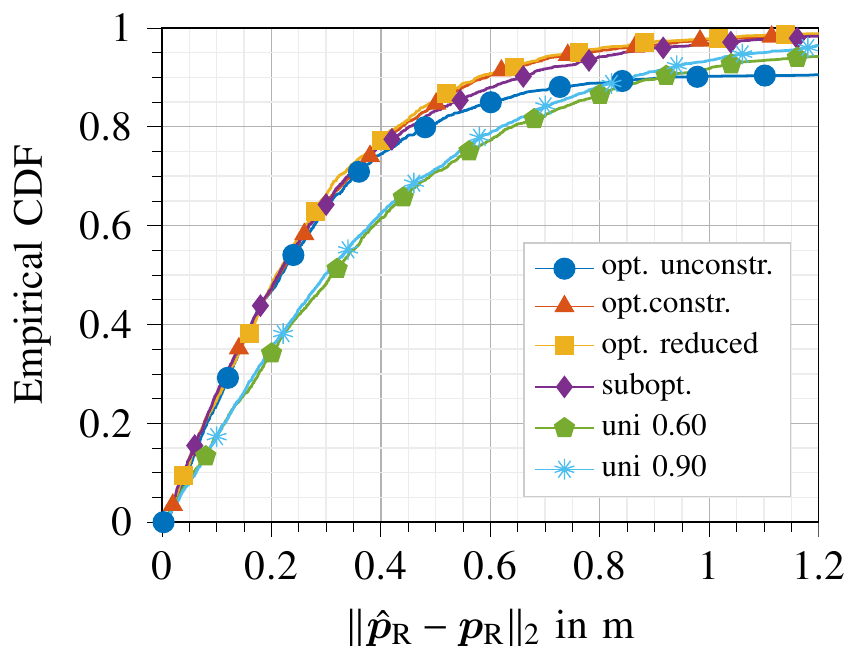}
			\caption{Empirical \gls{cdf} of $\| \hat{\vect{p}}_{\text{R}} - \vect{p}_{\text{R}} \|_2$ for different power allocation strategies.}
			\label{fig:position_error_cdf}
		\end{figure}

		The results in Fig.~\ref{fig:position_RMSE_and_RCRLB_vs_tx_power} and the corresponding discussion were useful in examining the behavior of the position estimation algorithm, but do not provide a complete characterization of the performance of the power allocation strategies. To better evaluate their performance, for $P_{\text{RE}} = \SI{0}{\dBm}$ and the rest of the system parameters as described in Sec.~\ref{sec:numerical results, system parameters}, we plot in Fig.~\ref{fig:position_error_cdf} the \gls{cdf} of the position error $\| \hat{\vect{p}_{\text{R}}} - \vect{p}_{\text{R}} \|_2$, which is computed drawing samples from~\eqref{eq:numerical results mean}-\eqref{eq:numerical results covariance}.
		A summary of the percentiles of the distribution of the position error is provided in Table~\ref{tab:percentiles}.
		\begin{table}
			\centering
			\caption{Percentiles of the \gls{cdf} of the position error in \si{\meter} for different power allocation strategies.}
			\setlength{\extrarowheight}{1pt}
			\begin{tabular}{r|cccc}
				& 50\% 	& 90\% & 95\% 	& 99\% \\
				\hline\hline
				opt. unconstr. 	& 0.22 	& 0.93 & 29.55 	& 72.25 \\
				opt. constr. 	& 0.21 	& 0.59 & 0.78 	& 1.32 \\
				opt. reduced 	& 0.21 	& 0.57 & 0.76 	& 1.25 \\
				subopt. 		& 0.21 	& 0.65 & 0.84 	& 1.45 \\
				uni 0.60 		& 0.31 	& 0.91 & 1.30 	& 20.83 \\
				uni 0.90 		& 0.30 	& 0.86 & 1.10 	& 1.96 \\
			\end{tabular}
			\label{tab:percentiles}
		\end{table}
		
		We can observe in Fig.~\ref{fig:position_error_cdf} and Table~\ref{tab:percentiles} that "opt. reduced" and "opt. constr." achieve the best performance. The latter is slightly worse at higher percentiles, as more points would be required for a more accurate approximation of the expectation in the corresponding optimization problem.
		In spite of the lower computations cost of the "subopt." allocation, its performance degradation is almost unnoticeable. 
		On the other hand, the "opt. unconstr." approach, although attaining almost the same median error as the other optimized strategies, has much lower accuracy for higher percentiles. This is attributed to the low power spent in the direction around the \gls{LOS} path, resulting in low probability of detection of the \gls{LOS}.
		Compared to the best of the uniform allocations, the "opt. reduced" power allocation offers a position error reduction of 30\%, 34\%, 31\% and 36\% at the 50\%, 90\%, 95\% and 99\% percentile, respectively.
		
		Regarding the uniform allocations, we can see that spreading the power to a reduced set of beams ("uni 0.60") might result in better positioning accuracy for some geometry realizations, as seen for example in Fig.~\ref{fig:position_RMSE_and_RCRLB_vs_tx_power}, but it significantly deteriorates the performance for other possible realizations. This explains the higher values of position errors at the upper percentiles of the corresponding \gls{cdf}.

		
		\subsection{Power allocation as a function of $\sigma_{\text{clk}}$}
\begin{figure}
			\centering
			\subfloat[$q_{\text{LOS}}$ vs. $\sigma_{\text{clk}}$]{\begin{adjustbox}{scale=0.97}
					\includegraphics[scale=0.90]{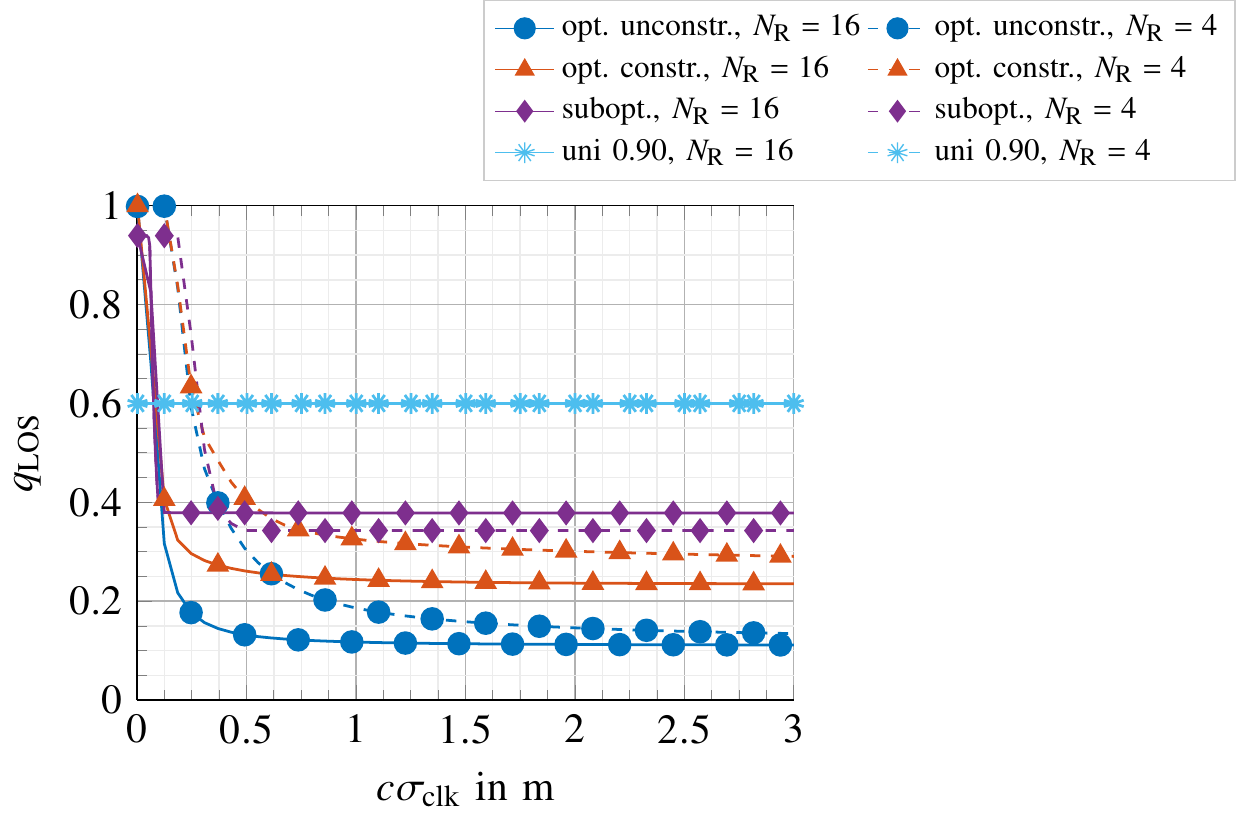}
			\end{adjustbox}}%
			\hspace*{-4cm}
			\subfloat[$\mathbb{E}{[\text{PEB}]}$ vs. $\sigma_{\text{clk}}$]{\begin{adjustbox}{scale=0.97}
					\includegraphics[scale=0.90]{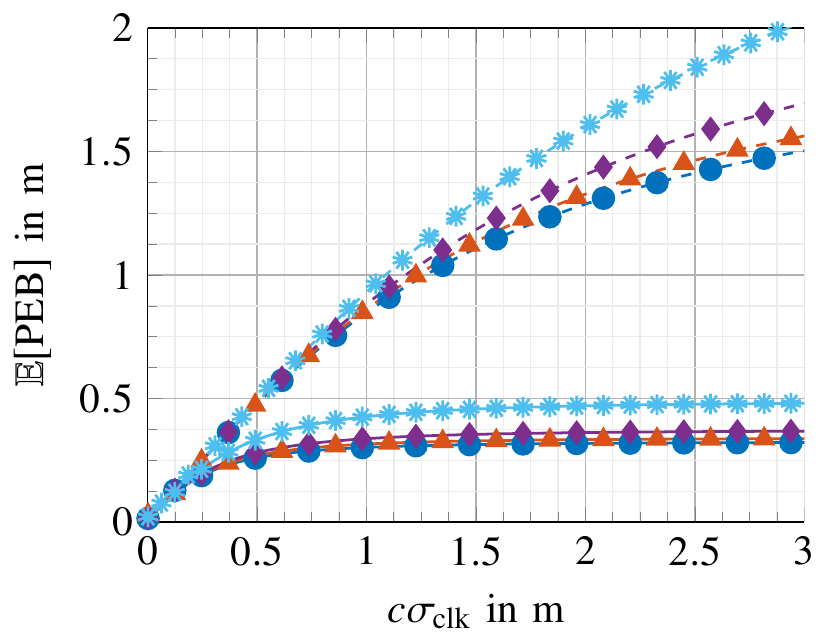}
			\end{adjustbox}}
			\caption{Fraction of power allocated to \gls{LOS}-illuminating beams $q_{\text{LOS}}$ and $\mathbb{E}[\text{PEB}]$ as functions of $\sigma_{\text{clk}}$.}
			\label{fig:power_to_LOS_vs_sigma_clk}
		\end{figure}

		We now examine the effect of $\sigma_{\text{clk}}$ on the power allocation. First, similar to \eqref{eq:useful beams}, we define the set of LOS-illuminating beams as
		\begin{IEEEeqnarray}{rCl}
			\mathcal{B}_{\text{LOS}}^{(\kappa)} &=& \cup_{m=0}^{N_{\theta}} \Big\{\argmax_{k=1,\ldots,N_{\text{T}}}|\vect{a}_{\text{T}}\trans(\theta_{\text{T}, 0, m}^{(\kappa)})\vect{f}_k|\Big\}. 
			\label{eq:LOS beams}
		\end{IEEEeqnarray}  
		and the fraction of power spent on them as
		\begin{IEEEeqnarray}{rCl}
			q_{\text{LOS}} &=& \sum_{k\in\mathcal{B}_{\text{LOS}}^{(\kappa)}} q_k.
		\end{IEEEeqnarray}
		In Fig.~\ref{fig:power_to_LOS_vs_sigma_clk}(a) we plot $q_{\text{LOS}}$ as a function of $\sigma_{\text{clk}}$ for the power allocation strategies "opt. unconstr.", "opt. constr.", "subopt" and "uni 0.90", for $N_{\text{R}} = \{4,\;16\}$, $P_{\text{RE}} = \SI{0}{\dBm}$, $\kappa=0.995$ and the rest of the system parameters as described in Sec.~\ref{sec:numerical results, system parameters}; in Fig.~\ref{fig:power_to_LOS_vs_sigma_clk}(b) we plot the corresponding $\mathbb{E}[\text{PEB}]$.
		We can see in Fig.~\ref{fig:power_to_LOS_vs_sigma_clk}(a) that for very low values of $\sigma_{\text{clk}}$, equivalent to almost perfect \gls{Tx}-\gls{Rx} synchronization, it is optimal to spend almost all the available power on LOS-illuminating beams. As $\sigma_{\text{clk}}$ increases, $q_{\text{LOS}}$ decreases rapidly for both optimized allocation strategies, until it saturates at a relatively low value. This is explained as follows: The clock offset decreases the amount of range information provided by the \gls{LOS} path and the larger standard deviation of the clock offset, the more significant the decrease. Hence, as $\sigma_{\text{clk}}$ increases, the ranging information provided by the NLOS paths becomes more significant and, therefore, more power is spent on them. Nevertheless, the saturation occurs because the measurement of the \gls{LOS} \gls{AOD} offers significant information in the orthogonal direction, which is reduced when $q_{\text{LOS}}$ is decreased. The saturation value for "opt. constr." is higher due to the additional constraints on LOS illumination. Also, we observe that the transition from high to low $q_{\text{LOS}}$ values is slower for $N_{\text{R}} = 4$.
		This is attributed to the fact that \gls{NLOS} paths offer rank-1 position information, whose intensity depends on the quality of the \gls{TOA}, \gls{AOD} and \gls{AOA} measurements combined~\cite{KCS+19}. For $N_{\text{R}} = 4$, the quality of the \gls{AOA} measurement is lower; therefore, the intensity of the ranging information from the \gls{NLOS} paths is smaller, compared to $N_{\text{R}} = 16$, and becomes significant for larger values of $\sigma_{\text{clk}}$.
		
		In Fig.~\ref{fig:power_to_LOS_vs_sigma_clk}(b) we see that $\mathbb{E}[\text{PEB}]$ increases with increasing $\sigma_{\text{clk}}$, until it saturates at a value dependent on the power allocation strategy and the system configuration ($N_{\text{R}} = \{4,\;16\}$). As $\sigma_{\text{clk}}$ increases the reduction of ranging information from the \gls{LOS} path cannot be complemented by ranging information from the NLOS paths (even with optimized power allocation), resulting in a larger error. 
		In the saturation region the ranging information from the \gls{LOS} path becomes negligible compared to the clock offset-independent part of ranging information offered by the combination of NLOS paths with the LOS path.
		
	\section{Conclusion}
		\label{sec:conclusion}
		Optimal power allocation on a beam codebook for single-anchor localization and lower-complexity suboptimal alternatives have been considered under imperfect \gls{Tx}-\gls{Rx} synchronization. A channel and position estimation method has also been proposed. Numerical results show that our suboptimal power allocation approach offers a good balance between performance and complexity, as the significant complexity reduction for the computation of the power allocation incurs only a very small performance penalty. Our analysis has shown that even for low values of the clock offset standard deviation it is optimal according to the \gls{CRLB} to allocate most of the available power to scatterer/reflector illuminating beams to recover necessary range information. We have also shown that guaranteeing a minimum amount of power spent on \gls{LOS}-illuminating beams, can be beneficial when the actual position estimation is considered, as it ensures that \gls{LOS} path is detected with a high probability. The proposed position estimation algorithm attains the corresponding \gls{CRLB} for all considered power allocation strategies, benefiting from the gridless parameter estimation, which avoids the appearance of spurious paths due to grid mismatch, while also filtering out noisy detected paths exploiting the information on the clock offset carried by single-bounce-\gls{NLOS} paths.
	
	
	\appendix[Power Allocation for the LOS Path]
		\label{sec:power allocation for the LOS path}
		Here we show how to formulate~\eqref{eq:suboptimal LOS} as an \gls{SDP} using only a \gls{1D} quadrature rule for the approximation of the expectation over $\theta_{\text{T}, 0}$. This is accomplished in two steps: 
		\begin{itemize}
			\item In the first step we show that the integration over $d_0$ and $\theta_{\text{T}, 0}$ can be carried out separately;
			\item in the second step, after averaging over $d_0$, we exploit the form of the resulting function of $\theta_{\text{T}, 0}$ and formulate the problem as an \gls{SDP}.
		\end{itemize}We write $\mathbb{E}_{d_0, \theta_{\text{T}, 0}}[\cdot]$ instead of $\mathbb{E}_{\vect{p}_{\text{R}}}[\cdot]$.
		Also, for notational brevity we write 
		\begin{IEEEeqnarray}{rCl}
			\bar{\vect{J}} = \mathbb{E}_{\alpha_{\text{R}}, \vect{h}_0|d_0, \theta_{\text{T}, 0}}[\vect{J}_{\vect{\nu}_{\text{LOS}}}(\vect{q}, d_0, \theta_{\text{T}, 0}, \alpha_{\text{R}}, \vect{h}_0)].
		\end{IEEEeqnarray}
		We index the elements of $\bar{\vect{J}}$ with the pair of parameters to which they correspond. 
		
		First, after some algebra we find that
		\begin{IEEEeqnarray}{rCl}
			\trace(\vect{E}\trans \bar{\vect{J}}^{-1} \vect{E} )&=&  \frac{c^2}{\bar{J}_{\tau_0,\tau_0} - \frac{\bar{J}_{\tau_0,\theta_{\text{T}, 0}}^2}{\bar{J}_{\theta_{\text{T}, 0}, \theta_{\text{T}, 0}}}}  + \frac{d_0^2}{\bar{J}_{\theta_{\text{T}, 0}, \theta_{\text{T}, 0}} - \frac{\bar{J}_{\tau_0,\theta_{\text{T}, 0}}^2}{\bar{J}_{\tau_0,\tau_0}}} + c^2\sigma_{\text{clk}}^2,\label{eq:LOS SPEB}
		\end{IEEEeqnarray}
		where
		\begin{IEEEeqnarray}{rCl}
			\bar{J}_{a,b} &=& \mathbb{E}_{\alpha_{\text{R}}, \vect{h}_0|d_0, \theta_{\text{T}, 0}}[J_{a,b}]\\
			J_{a,b} &=& \frac{2}{\sigma_{\eta}^2}\sum_{b=1}^{N_{\text{B}}} \sum_{p\in \mathcal{P}} \Re\bigg\{\frac{\partial \vect{m}_b\herm[p]}{\partial a} \frac{\partial \vect{m}_b[p]}{\partial b}\bigg\},
			\IEEEeqnarraynumspace
		\end{IEEEeqnarray}
		with $a,b\in\{d_0,\theta_{\text{T}, 0}\}$.
		We can show that $J_{a,b},\;a,b\in\{d_0, \theta_{\text{T}, 0}\}$, are independent of $\alpha_{\text{R}}$ and the phase of $h_0$. 
		Hence, they can be expressed as 
		\begin{IEEEeqnarray}{rCl}
			\bar{J}_{a,b} &=& \mathbb{E}_{\vect{h}_0|d_0, \theta_{\text{T}, 0}} [J_{a,b}(\vect{q}, \theta_{\text{T}, 0}, |h_0(d_0)|^2)]=\mathbb{E}_{\vect{h}_0|d_0, \theta_{\text{T}, 0}} [|h_0(d_0)|^2 j_{a,b}(\vect{q}, \theta_{\text{T}, 0})]=g_0(d_0) j_{a,b}(\vect{q}, \theta_{\text{T}, 0}),\label{eq:barJ_ab}
			\IEEEeqnarraynumspace
		\end{IEEEeqnarray}
		where $g_0(d_0) = \mathbb{E}_{\vect{h}_0|d_0} [|h_0(d_0)|^2]$ and $j_{a,b}(\vect{q}, \theta_{\text{T}, 0}) = J_{a,b}(\vect{q}, \theta_{\text{T}, 0}, |h_0(d_0)|^2)/|h_0(d_0)|^2$ is a function of $\vect{q}$ and $\theta_{\text{T}, 0}$. 
		For the second equality in~\eqref{eq:barJ_ab}, 
		we used the fact that $J_{a,b}$ can be expressed as the product of two terms, 
		one dependent on the gain magnitude and the other on $\vect{q}$ and $\theta_{\text{T}, 0}$. 
		We can then rewrite \eqref{eq:LOS SPEB} as 
		\begin{IEEEeqnarray}{rCl}
			\trace(\vect{E}\trans \hspace{-0.05cm} \bar{\vect{J}}^{-1}\hspace{-0.05cm} \vect{E} )\hspace{-0.03cm}&=&\hspace{-0.03cm} \frac{1}{g_0(d_0)} \bigg(\frac{c^2}{I_{\tau_0}\hspace{-0.01cm}(\hspace{-0.01cm}\vect{q},\hspace{-0.01cm}\theta_{\text{T}, 0}\hspace{-0.01cm})}\hspace{-0.03cm} + \hspace{-0.03cm}\frac{d_0^2}{I_{\theta_{\text{T}, 0}}\hspace{-0.01cm}(\hspace{-0.01cm}\vect{q},\hspace{-0.01cm}\theta_{\text{T}, 0}\hspace{-0.01cm})}\bigg)\hspace{-0.03cm} +\hspace{-0.03cm} c^2\sigma_{\text{clk}}^2,\label{eq:LOS SPEB 2}
			\IEEEeqnarraynumspace
		\end{IEEEeqnarray}
		where
		\begin{IEEEeqnarray}{rCl}
			I_{\tau_0}(\vect{q},\theta_{\text{T}, 0}) &=& j_{\tau_0,\tau_0}(\vect{q}, \theta_{\text{T}, 0}) - \frac{j_{\tau_0,\theta_{\text{T}, 0}}^2(\vect{q}, \theta_{\text{T}, 0})}{j_{\theta_{\text{T}, 0}, \theta_{\text{T}, 0}}(\vect{q}, \theta_{\text{T}, 0})}\\
			I_{\theta_{\text{T}, 0}}(\vect{q},\theta_{\text{T}, 0}) &=& j_{\theta_{\text{T}, 0}, \theta_{\text{T}, 0}}(\vect{q}, \theta_{\text{T}, 0}) - \frac{j_{\tau_0,\theta_{\text{T}, 0}}^2(\vect{q}, \theta_{\text{T}, 0})}{j_{\tau_0,\tau_0}(\vect{q}, \theta_{\text{T}, 0})}.
		\end{IEEEeqnarray}
		It is then apparent from the form of the function in \eqref{eq:LOS SPEB 2} that integration over $d_0$ and $\theta_{\text{T}, 0}$ can be carried out separately. 
		
		For the second step, taking the expectation over $d_0$ and defining
		\begin{IEEEeqnarray}{rCl}
			\bar{g}_0(\theta_{\text{T}, 0}) &=& 1/\mathbb{E}_{d_0|\theta_{\text{T}, 0}}[1/g_0(d_0)]\\
			\bar{d}_0(\theta_{\text{T}, 0}) &=& \sqrt{\mathbb{E}_{d_0|\theta_{\text{T}, 0}}\bigg[\frac{\bar{g}_0(\theta_{\text{T}, 0})}{g_0(d_0)} d_0^2\bigg]}
		\end{IEEEeqnarray}
		we get
		\begin{IEEEeqnarray}{rCl}
			\mathbb{E}_{d_0|\theta_{\text{T}, 0}}[\trace(\vect{E}\trans \hspace{-0.05cm} \bar{\vect{J}}^{-1}\hspace{-0.05cm} \vect{E} )\hspace{-0.03cm}]&=&\hspace{-0.03cm} \frac{1}{\bar{g}_0(\theta_{\text{T}, 0})} \bigg(\frac{c^2}{I_{\tau_0}\hspace{-0.01cm}(\hspace{-0.01cm}\vect{q},\hspace{-0.01cm}\theta_{\text{T}, 0}\hspace{-0.01cm})}\hspace{-0.03cm} + \hspace{-0.03cm}\frac{\big(\bar{d}_0(\theta_{\text{T}, 0})\big)^2}{I_{\theta_{\text{T}, 0}}\hspace{-0.01cm}(\hspace{-0.01cm}\vect{q},\hspace{-0.01cm}\theta_{\text{T}, 0}\hspace{-0.01cm})}\bigg)\hspace{-0.03cm} + c^2\sigma_{\text{clk}}^2\label{eq:LOS SPEB 3}.
			\IEEEeqnarraynumspace
		\end{IEEEeqnarray}
		Comparing \eqref{eq:LOS SPEB 3} to \eqref{eq:LOS SPEB}, 
		we can conclude that, in order to be able to formulate the problem in a convex form, 
		$\mathbb{E}_{d_0|\theta_{\text{T}, 0}}[\trace(\vect{E}\trans \hspace{-0.05cm} \bar{\vect{J}}^{-1}\hspace{-0.05cm} \vect{E} )\hspace{-0.03cm}]$ can be expressed as 
		\begin{IEEEeqnarray}{rCl}
			\mathbb{E}_{d_0|\theta_{\text{T}, 0}}&&[\trace(\vect{E}\trans \hspace{-0.05cm} \check{\vect{J}}^{-1}\hspace{-0.05cm} \vect{E} )\hspace{-0.03cm}]= \trace(\vect{E}\trans \vect{J}_{\vect{\nu}_{\text{LOS}}}^{-1} (\vect{q}, \bar{d}_0(\theta_{\text{T}, 0}), \theta_{\text{T}, 0}, \check{\alpha}_{\text{R}}, \sqrt{\bar{g}_0(\theta_{\text{T}, 0})}\e^{\jj \beta_g})\vect{E}), \label{eq:LOS SPEB 4}
			\IEEEeqnarraynumspace
		\end{IEEEeqnarray}
		where $\check{\alpha}_{\text{R}}$ and $\beta_g$ can be chosen arbitrarily, since they do not have an impact on the objective.
		Finally, using \eqref{eq:LOS SPEB 4} and the identity
		\begin{IEEEeqnarray}{rCl}
			\mathbb{E}_{d_0, \theta_{\text{T}, 0}} \big[ \trace(\vect{E}\trans \bar{\vect{J}} \vect{E} ) \big] &=& \mathbb{E}_{\theta_{\text{T}, 0}}[\mathbb{E}_{d_0|\theta_{\text{T}, 0}}[\trace(\vect{E}\trans \hspace{-0.05cm} \bar{\vect{J}}^{-1}\hspace{-0.05cm} \vect{E} )\hspace{-0.03cm}]],
		\end{IEEEeqnarray}
		we can employ a \gls{1D} quadrature rule to approximate the expectation integral over $\theta_{\text{T}, 0}$ to get the following \gls{SDP}:
		\begin{IEEEeqnarray}{rCl}
			&&\min_{\vect{q}, \vect{B}_1, \ldots, \vect{B}_{N_{\theta_{\text{T}, 0}}}} \sum\nolimits_{j=1}^{N_{\theta_{\text{T}, 0}}} p_j \trace(\vect{B}_j)\nonumber\\
			\text{s.t. }&&\vect{q}\geq \vect{0},\; \vect{1}\trans\vect{q}\leq 1,\nonumber\\
			&& \begin{bmatrix}
				\vect{B}_j & \vect{E}\trans\\
				\vect{E} & \vect{J}_{\vect{\nu}_{\text{LOS}}}(\vect{q}, \bar{d}_0(\theta_{\text{T}, 0, j}), \theta_{\text{T}, 0, j}, \check{\alpha}_{\text{R}}, \sqrt{\bar{g}_0(\theta_{\text{T}, 0, j})}\e^{\jj \beta_g})\\
			\end{bmatrix}\succeq \vect{0}, \quad j=1,\ldots, N_{\theta_{\text{T}, 0}}.
			\label{eq:LOS SPEB SDP}
			\IEEEeqnarraynumspace
		\end{IEEEeqnarray}

		\bibliographystyle{IEEEtran}
		\bibliography{IEEEabrv,References}
	
\end{document}